\documentclass[preprint,showpacs,preprintnumbers,amsmath,amssymb]{revtex4-1}

\usepackage[usenames,dvipsnames]{xcolor}
\usepackage[T1]{fontenc}
\usepackage{graphicx}
\usepackage[caption=false,margin=0pt]{subfig}
\usepackage[utf8]{inputenc}
\usepackage{textcomp}
\usepackage{graphicx}
\usepackage{braket}
\usepackage{multirow}

\usepackage[colorlinks=true,linkcolor=blue,citecolor=blue,urlcolor=blue]{hyperref}%

\newcommand{\Avec}{\vec{A}}
\newcommand{\epsh}{\hat\varepsilon}

\begin{document}
\title{Tuning the Fermi surface of MoTe$_2$ by Nb doping}

\author{Andrew P. Weber$^{1,2}$}
\author{Iñigo Robredo$^{3}$}
\author{Philipp Rü{\ss}mann$^{4}$}
\author{Maxim Ilyn$^{5}$}
\author{Arnaud Magrez$^{6}$}
\author{Philippe Bugnon$^{6}$}
\author{Nan Xu$^{7,8}$}
\author{Vladimir N. Strocov$^{9}$}
\author{J. Hugo Dil$^{6,9}$}
\author{J. Enrique Ortega$^{5,10}$}
\author{Julen Ibañez-Azpiroz$^{1,5,11}$}
\email{julen.ibanez@ehu.eus}

\affiliation{$^{1}$Donostia International Physics Center, 20018 Donostia, Gipuzkoa, Spain}
\affiliation{$^{2}$ICFO-Institut de Ciencies Fotoniques, The Barcelona Institute of Science and Technology, 08860 Castelldefels (Barcelona), Spain}
\affiliation{$^{3}$Smart Materials Unit, Luxembourg Institute of Science and Technology (LIST), Avenue des Hauts-Fourneaux 5, L-4362 Esch/Alzette, Luxembourg}
\affiliation{$^{4}$Peter Grünberg Institut and Institute for Advanced Simulation, Forschungszentrum Jülich and JARA, 52425 Jülich, Germany}
\affiliation{$^{5}$Centro de Física de Materiales CSIC-UPV/EHU and Materials Physics Center, 20018 San Sebastián, Spain}
\affiliation{$^{6}$Institute of Physics, \'{E}cole Polytechnique F\'{e}d\'{e}rale de Lausanne, CH-1015, Lausanne, Switzerland}
\affiliation{$^{7}$ Institute for Advanced Studies, Wuhan University, Wuhan 430072, China} 
 \affiliation{$^{8}$ Wuhan Institute of Quantum Technology, Wuhan 430206, China}
\affiliation{$^{9}$Center for Photon Science, Paul Scherrer Institute, CH-5232 Villigen, Switzerland}
\affiliation{$^{10}$Departamento de Física Aplicada, Universidad del País Vasco, 20018 San Sebastián, Spain}
\affiliation{$^{11}$IKERBASQUE, Basque Foundation for Science, 48009 Bilbao, Spain}

\date{\today}
\begin{abstract}
\textit{Ab initio} calculations and angle-resolved photoemission experiments {demonstrate} that the bulk and surface electronic structure of Weyl semimetal candidate MoTe$_2$ changes significantly by tuning the chemical potential by less than 0.4 eV. Calculations {indicate} that several Lifshitz transitions can occur among multiple electron and hole Fermi pockets of differing orbital character. Experiments show that 18\% Nb-Mo substitution reduces the occupation of bulk and (001) surface bands, effectively producing a {quasi-rigid} band shift 
of $\approx 0.2$~eV. Orbital character and dimensionality of the bulk bands is examined by soft X-ray angle resolved photoemission with control of the excitation light polarization. Deposition of alkali atoms on the surface produces electron binding energy shifts consistent with increased band filling. The results indicate that multiple regimes of electronic properties can be easily accessed in this versatile, layered material.

\end{abstract}


\maketitle

\section{Introduction}

Transition metal dichalcogenides based on (Mo/W)Te$_2$ are versatile layered materials hosting a range of desirable and tune-able properties. Topological insulator \cite{Shi2019} and semimetal \cite{Tamai2016} phases, exotic magnetotransport \cite{Ivo2014,Keum2015,Yang-Yang2017,Wu2018,Liang2019,Li2019} and optical \cite{Herrero2018,PhysRevB.108.165418} behavior, and attractive catalytic \cite{McGlynn2019} and thermoelectric \cite{Sakai2016,Takahashi2019} properties have been found in the various structural polytypes. The hexagonal 2H structure is semiconducting whereas monoclinic 1T'-MoTe$_2$ is semimetallic \cite{Brown1966} and undergoes a transition upon cooling below 250 K to the orthorhombic T$_d$, which hosts the Weyl semimetal phase \cite{Tamai2016,Jiang2017,Deng2016,Huang2016,Sakano2017,Ruessmann2018,Berger2018,Weber2018,Xu2018} and may be a topological superconductor below 1~K \cite{Qi2016,Luo2016,Takahashi2017,Guguchia2017,Mandal2018,Cui2019,Gan2019}. 

The polytypes can be controllably interchanged by virtue of the small differences in energy between structures \cite{Kim2017}. For instance, the 2H phase can be changed to the 1T' phase by electrostatic doping \cite{Wang2017,Li2016,Zhang2019}, strain \cite{Song2016,Hou2019}, laser irradiation \cite{Tan2018}, and high electric fields \cite{Kim2020}, while subpicosecond transition to the 1T' phase can be optically induced in T$_d$-MoTe$_2$ \cite{Zhang2019subpico}. Reduced dimensionality \cite{He2018,Cao2018}, interfaces \cite{Tsipas2018,Weber2018,Huang2019motedomains}, and pressure \cite{Heikes2018,Takahashi2017,Takahashi2019} can also have a strong effect on the relative stability of the 1T' and T$_d$ phases. The T$_d$ phase is observed at room temperature in ultrathin 1T'-MoTe$_2$ \cite{He2018,Cao2018,Tsipas2018}, while incoherent T$_d$ order persists near the surface of bulk 1T' crystals at room temperature \cite{Weber2018} and volume fractions of both phases exist within bulk crystals \cite{Clarke1978,Chen2016nano,Sakai2016,Yan2017a,Heikes2018} and coherent intermediate structures~\cite{Tao2019}.

Partial chemical substitution and Te vacancies are an efficient way of modifying the properties of MoTe$_2$. For instance, the electron-doping effect of Te vacancies is related to the presence of superconductivity below 1 K, suppression of longitudinal magnetoresistance (MR), and broadening of the 1T'/T$_d$ transition \cite{Cho2017}. The vacancies also play an important role in non-reversible 2H/1T' phase transitions \cite{Tan2018,Zhang2019,Kim2020}, and elemental substitution can be used to vary the polytype obtained with given reaction conditions \cite{Jin2018,Rhodes2017MoWTe2,Li2019moTeSe,Ikeura2015,Sakai2016,Kang2018}. 

Furthermore, the relative energy of the 1T' and T$_d$ phase is very sensitive to the occupation of low-lying energy bands, offering the opportunity to tune the structural ground state and transition energy barrier \textit{via} mild charge doping \cite{Kim2017}. In practice, effective hole doping can be achieved by substitution of Mo with Nb, reducing the $d$ electron count \cite{Ikeura2015,Kang2018}. Previous studies in 1T'-Mo$_{1-x}$Nb$_{x}$Te$_2$ alloys have shown that the T$_d$ phase is suppressed, 
leading to short-ranged polar ordering accompanied by enhanced thermopower \cite{Sakai2016}.

In this work, we characterize the effect of Nb-Mo substitution and surface alkali doping on the occupied electronic structure of 1T'-MoTe$_2$ by angle resolved photoemission (ARPES), and we use soft X-ray ARPES (SX-ARPES) as a probe of three-dimensional (3D) electronic band structure \cite{Xu2018}.
We combine the experimental measurements with 
density functional theory (DFT) calculations  
in order to visualize the expected response to changes in chemical potential $\mu_F$ and the differences in the wave-vector-dependent orbital composition (orbital texture) in the bands.
We find that the electron Fermi pockets that are present in pristine bulk MoTe$_2$ disappear upon Nb-Mo substitution, which lowers the Fermi energy well into the hole-like valence bands. The dispersion and constant energy contour shapes of the bands are slightly altered and 3D electronic coherence is retained. 
Furthermore, our study based on ultraviolet ARPES (UV-ARPES) reveals that significant electron doping can be achieved by depositing alkali atoms on the surface of Mo$_{0.82}$Nb$_{0.18}$Te$_{1.91}$. Overall, our results show that 
the combination of Nb substitution and surface doping can effectively tune the chemical potential, allowing  the tuning of conduction properties.   

The paper is organized as follows. In Sec. \ref{sec:methods} we describe technical aspects concerning the ARPES measurements and DFT calculations  
reported throughout the work. 
In Sec. \ref{sec:symmetry} we analyze the symmetry properties 
and expected implications on the theoretical and experimental results.  
Sec. \ref{sec:results} then contains the  bulk of our results; in Sec. \ref{subsec:dft} we provide a detailed theoretical analysis
of ground state properties of the bulk and surface, 
while Sec. \ref{subsec:exp} contains the experimental characterization of the system
based on various types of ARPES measurements performed in 
samples with and without Nb doping. 
Finally,  we summarize our main results in 
Sec. \ref{sec:summary}.
Technical details are kept for Appendices~\ref{sec:App-A} and~\ref{sec:App-B}.

\section{Methods}
\label{sec:methods}

\subsection{ARPES measurements}

Single-crystals of 1T'-MoTe$_{1.96}$ and 1T'-Mo$_{0.82}$Nb$_{0.18}$Te$_{1.91}$ were used 
in the experimental studies and are henceforth referred to as $x=0$ and $x=0.18$ samples, 
respectively. 
Synthesis and characterization of the $x=0$ samples was reported previously \cite{Ruessmann2018}. 
The $x=0.18$ samples were prepared by chemical vapor transport. 
High purity Mo, Te and Nb metals were sealed in quartz ampule together with iodine used as transport agent. The nominal molar stoichiometry of Mo/Nb/Te was 0.8/0.2/2. Optimum crystal growth temperatures were found to be 1000 $^{\circ}$C and 940 $^{\circ}$C at the source and sink, respectively. 
The Mo/Nb/Te stoichiometry in the resulting crystals was determined to be 0.82(2)/0.18(2)/1.9(1) by energy-dispersive X-ray spectroscopy. 
Single-crystal X-ray diffraction was measured with the sample kept at a temperature of $T=80$ K. 
The $P2_{1}/m$ (No. 11) structure was confirmed and the refined lattice parameters were found to be $a = 6.395(1)$, $b = 3.504(1)$, $c = 13.496(5)$,  $\beta = 92.64(1)^{\circ}$.
Temperature-dependent magnetotransport measurements, shown in Appendix~\ref{sec:App-B}, were consistent with a previous report \cite{Sakai2016} and showed the absence of a sharp transition to the T$_d$ phase for $x=0.18$.

ARPES experiments were performed with the sample kept under ultrahigh vacuum (UHV) (pressure $<1\times10^{-9}$ Pa) at variable temperatures stated in the text. Clean (001) surfaces were obtained by cleaving in UHV. 
Soft X-ray ARPES (SX-ARPES) measurements were performed 
{using the SX-ARPES endstation~\cite{Strocov_ve5018}}
at the Advanced Resonant Spectroscopies beam line at the Swiss Light Source \cite{Strocov2010}.
UV-ARPES of undoped MoTe$_2$ was performed at beamline X09LA of the Swiss Light Source. 
The UV-ARPES study of surface Rb-doping was carried out in a laboratory based ARPES setup using 21.22 eV photons from a He discharge lamp.  
The total energy and angular resolution was better than 90 meV and 0.1\textdegree\ for SX-ARPES and better than 35 meV and 0.4\textdegree\ for UV-ARPES measurements. 
A commercial SAES Rb dispenser was used for \textit{in situ} surface doping. 
A quartz crystal microbalance was used to estimate the rate of the deposition, which was performed and studied with the sample kept at 60 K.

\subsection{Theoretical calculations}

While the pristine 1T'-MoTe$_2$ structure is centrosymmetric, 
the inversion symmetry of the crystal breaks upon Nb substitution. 
Among other features, 
this causes  a general splitting of originally spin-degenerate bands
through the action of spin-orbit coupling.
In order to capture this and further effects, we have focused our theoretical
analysis on the T$_d$-MoTe$_2$ structure,
given that this polytype is naturally acentric and its symmetry properties are
therefore more in line with the experimentally measured system. 
This choice is further justified by the fact that the electronic band dispersion, metallic character and  
density of states near the Fermi level
of 
the 1T' and T$_d$ polytypes are very similar, as well as their calculated total energy~\cite{Crepaldi2017,Sankar2017,hernandez_ruiz_chemical_2022}.
Furthermore, previous experimental studies on MoTe$_2$~\cite{Ikeura2015,Sakai2016} and theoretical calculations for the related compound WSe$_2$~\cite{Hisama_2022} have shown that Nb doping is well described by a rigid-band shift, with only minor changes to the underlying electronic structure; Ref. for instance estimated a rigid shift of 
approximately 125~meV for $x=0.1$ in Mo$_{1-x}$Nb$_x$Te$_2$~\cite{Ikeura2015}.
Moreover, the electronic structure is largely insensitive to both the dopant concentration and the specific dopant configuration~\cite{Hisama_2022}.
Motivated by these findings, we model the effect of Nb substitution in our calculations through variations of the Fermi level.

%

We have performed the DFT bulk calculations using the {\tt Quantum ESPRESSO} code
package~\cite{gianozzi-jpcm09}. 
We took the structural parameters for T$_d$-MoTe$_2$  from 
Ref.~\onlinecite{Qi2016}.
We treated the core-valence interaction 
by means of fully-relativistic projector augmented-wave
pseudopotentials  using the Perdew-Burke-Ernzerhof
exchange-correlation functional~\cite{perdew-prl96}, while 
we set the energy cutoff for the plane-wave basis expansion at 70~Ry.
We used $8\times 8\times 4$ \textit{k}-point mesh for the self-consistent calculation
and a $10\times 10\times 1$ one for the non self-consistent calculation employed for 
visualizing the Fermi surface cuts.
Finally, we computed the irreducible representations of the calculated Bloch states using the {\tt irrep} code~\cite{IRAOLA2022108226}
in combination with the {\tt Bilbao Crystallographic Server}~\cite{aroyo_bilbao_2006}.

In order to describe the surface-state dispersion measured in experiment, 
we have modeled a   semi-infinite T$_d$-MoTe$_2$ slab employing the local spin density approximation \citep{Vosko1980} and the full-potential relativistic Korringa-Kohn-Rostoker Green's function method (KKR) \citep{Ebert2011,jukkr} with exact description of the atomic cells \citep{Stefanou1990,Stefanou1991}. The truncation error arising from an $\ell_{max} = 3$ cutoff in the angular momentum expansion was corrected for using Lloyd's formula \citep{Zeller2004}.
To calculate the electronic structure of semi-infinite MoTe$_2$ we 
have used periodic boundary conditions in the plane parallel to the MoTe$_2$ layers and in the third direction we divide the space into three regions: \textit{A} the vacuum, \textit{B} the transition region describing the surface termination of MoTe$_2$ and \textit{C} the bulk continuation of MoTe$_2$. In region \textit{B} we have explicitly taken into account three MoTe$_2$ trilayers that are connected to the Green's function of the free space (region \textit{A}) and to the Green’s function of bulk MoTe$_2$ (region \textit{C}) using the principal layer technique \cite{Kudrnovsky1994}. The calculations for MoTe$_2$ were based on the thin film calculations used in Ref.~\cite{Weber2018}, where further numerical details can be found.
To visualize the band structures for semi-infinite MoTe$_2$, we 
have integrated the Bloch spectral function, which can be regarded as the $\vec{k}$-resolved density of states  over the layers of the transition region~\cite{Ebert2011}. This integral then includes the bulk-like states at the interface to region \textit{C} as well as the contribution of surface states at the termination to the vacuum (region \textit{A}).
To investigate the surface localization of the states, we have performed a partial integration over the layers of region \textit{B} to sum up contributions to the Bloch spectral function including the first MoTe$_2$ trilayer only. We have defined the ratio of this partial integration to the total integration over all layers of region \textit{B} as the surface localization of the states.

\section{Symmetry analysis of T$_d$-MoTe$_2$}
\label{sec:symmetry}
\subsection{Crystal Structure}

We first discuss several symmetry properties of bulk T$_d$-MoTe$_2$. Although these strictly apply only to the pristine T$_d$ structure, they nevertheless provide a useful approximate framework for interpreting our experimental results on the doped alloy in the 1T' polytype.

The crystal structure of T$_d$-MoTe$_2$ is depicted in Fig. \ref{fig:Td_struct}(a). It consists of identical Te$-$Mo$-$Te trilayers stacked along the $z$-direction, parallel to the crystallographic $c$-axis. Zigzag chains of Mo atoms, bridged by Te atoms labelled Te1 and Te3, are aligned along the $x$-direction, parallel to the crystallographic $a$-axis \cite{Titus2016}. The chains are coupled through bonding with the Te2 and Te4 atomic sites between them. The space group (No. 31) contains the mirror symmetry operation $M_x$, which takes a general position vector $(x,y,z)$ to $(-x,y,z)$, and the non-symmorphic glide-reflection $\widetilde{M}_y$, which takes $(x,y,z)$ to $(x+1/2,-y,z+1/2)$. The combination of these two gives rise to a 2-fold screw axis $C_{2_1}$ which takes $(x,y,z)$ to $(-x+1/2,-y,z+1/2)$. All of the atomic sites have $M_x$ symmetry, 
while $\widetilde{M}_y$ transforms the position of each atom into an equivalent sublattice and determines how the layers are stacked. Inversion symmetry is absent and does not exist locally in the individual MoTe$_2$ layers either.

\subsection{Constraints on Electronic Bands}
\label{subsec:symconstrains}

The breaking of inversion symmetry allows band-splittings in generic k-points. In turn, the presence of time-reversal symmetry (TRS) constrains the bands to become spin-degenerate at time-reversal invariant momentum (TRIM) points in the bulk Brillouin zone (BBZ) \cite{Ruessmann2018}. Further constraints on the band dispersion can be understood by considering the little co-group at different high symmetry points, lines and planes of the BBZ \citep{bradley1972mathematical, BCS1}, which is sketched in Fig. \ref{fig:Td_struct}(b). Red dashed-lines indicate how the high symmetry points of the BBZ project onto those of the surface Brillouin zone (SBZ) sketched above it, which will be discussed later. 

We focus our analysis on the $k_z=0$ and $k_z=\pi/c$ planes, and in particular on the $\Gamma$Z line, as these are central to the discussion of the 
band structure in Sec. \ref{subsec:dft}.
The combination of TRS and $C_{2_1}$ screw axis leaves both planes invariant. 
According to the representation content of their little co-group, there is only a two-dimensional irreducible representation at the $k_z=\pi/c$ plane, hence bands need to be at least two-fold degenerate there. 
In contrast, there are only one-dimensional irreducible representations in the little co-group at the $k_z=0$ plane, 
hence no symmetry-enforced degeneracy is imposed at this plane.
Along the  $\Gamma$Z line, the crystal symmetries $M_x$ and $\widetilde{M}_y$
do not commute and give rise to a two-dimensional irreducible representation, forcing bands to be doubly degenerate. 
As for the TRIMs $\Gamma$ and Z, their little co-group only contain  a two-dimensional and a four-dimensional irreducible representation, respectively; 
as a consequence, bands must be two-fold and four-fold degenerate at $\Gamma$ and Z, respectively.

\subsection{Glide-Reflection Eigenvalues}
\label{subsec:symm-evals}

\begin{figure}
\includegraphics[width=0.9\textwidth]{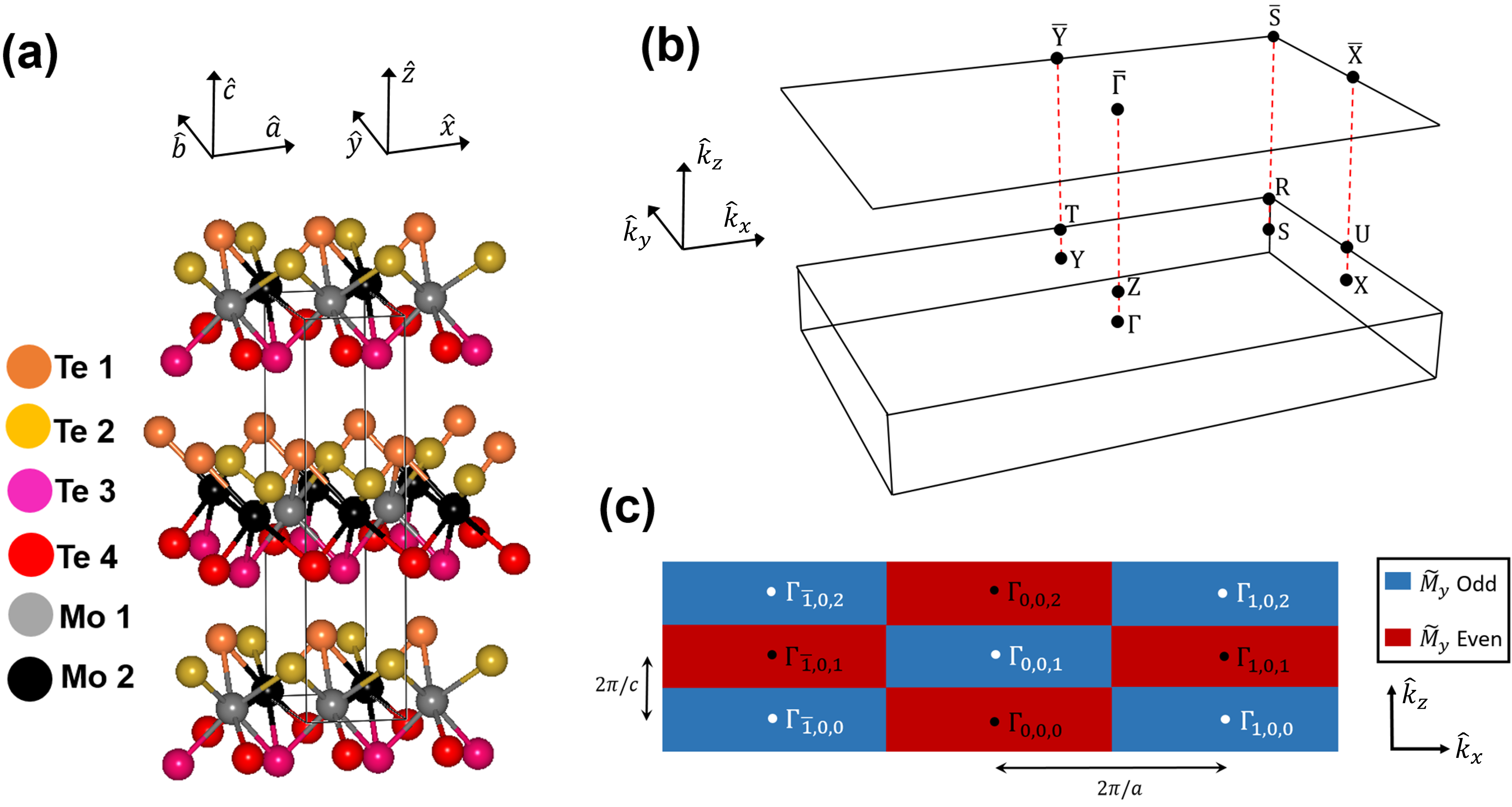}
\caption{(a) Crystal structure and (b) Brillouin zone of T$_d$-MoTe$_2$. The real space and momentum space axes are shown at the top of (a) and side of (b), respectively. Each set of equivalent sites is color-coded and labelled Te 1$-$Te 4 and Mo 1$-$Mo 2 as shown by the inset to the left. In (b), the bulk Brillouin zone is shown at the bottom with the corresponding (001) surface Brillouin zone. High symmetry momentum points are marked as dots and labelled. (c) Sketch of nine Brillouin zones in the $(k_x,k_z)$-plane, shaded red (blue) according to the even (odd) $\widetilde{M_y}$ parity of a plane wave state in that zone. The $\Gamma$ point at the center of each zone is labelled $\Gamma_{n_x,n_z}$ according to the $n_x$ and $n_z$ indices, where an overbar denotes a negative value.}
\label{fig:Td_struct}
\end{figure}

As a prerequisite for understanding the spin-orbital character discussed in Sec. \ref{subsec:dft} and ARPES results in Sec. \ref{subsec:exp}, the $\widetilde{M}_y$ eigenvalues for states in the $k_y=0$ plane of the BBZ are codified here. For this purpose, we describe the total wave-vector $\vec{k}=\vec{q}+\vec{G}$ as the sum of a vector $\vec{q}$ with its origin at $\Gamma_{n_x,n_y,n_z}$, the center of a BBZ indexed by the set of integers $n_x$, $n_y$, and $n_z$, and the reciprocal-lattice vector $\vec{G}$ between $\Gamma_{n_x,n_y,n_z}$ and the zero-momentum point $\Gamma_{0,0,0}$. The wave-function of a stationary state takes on the $\widetilde{M}_y$ eigenvalues $\pm e^{i(q_xa+q_zc)/2}$. Henceforth, the ``$+$'' and ``$-$'' cases are referred to as ``even'' and ``odd'' parity, respectively. The eigenvalue for a plane-wave, on the other hand, is $e^{i\pi(n_x+n_z)}e^{i(q_xa+q_zc)/2}$, such that the parity depends on the BBZ indices as illustrated in \ref{fig:Td_struct}(c). A checkerboard pattern is formed wherein the $\widetilde{M}_y$ eigenvalue alternates across each zone boundary perpendicular to the $k_x$ and $k_z$ axes. This has an effect on ARPES observations that is summarized here and described in more detail in Appendix~\ref{sec:App-A}. The photoemission current generated from a dipole-allowed transition between an initial, stationary state $\ket{i}$ and a plane-wave final state $\ket{f}$ is proportional to $|\bra{f}\vec{A}\cdot \hat{p}\ket{i}|^2$, where $\vec{A}$ is the vector potential and $\hat{p}$ is the momentum operator. Parity selection rules then dictate that the observation of a band with a given $\widetilde{M_y}$ depends on both the orientation of $\vec{A}$ \textit{and} the final state wave-vector, whose length along $k_z$ is a function of photon energy as shown in Eq. \ref{eq:kz} of Appendix~\ref{sec:App-A}. This relationship allows the three-dimensional band structure to be imaged by ARPES, but the selection rules cause the observed dispersion to have a periodicity twice that of the reciprocal lattice{~\cite{Pescia1985,Xu2018,PhysRevB.97.235122,PhysRevLett.119.026403,PhysRevX.7.041067}}. 

While the glide symmetry arguments discussed here strictly apply only to the $T_d$ phase, the presence of two MoTe$_2$ trilayers per unit cell in both $T_d$ and $1T'$ preserves the underlying photoelectron interference mechanism. Therefore, the resulting polarization dependence and $4\pi/c$ periodicity 
illustrated in Fig.~\ref{fig:Td_struct}(c)
remain a valid approximate framework for interpreting the doped alloy measurements on the
$1T'$ polytype.
In the course of the ARPES analysis in Sec.~\ref{subsec:exp}, it will be shown that variation of the light-polarization and photon energy can be used to exploit these effects for the purpose of resolving specific bands.  

\section{Results and Discussion}
\label{sec:results}
\subsection{DFT calculations of T$_d$-MoTe$_2$}
\label{subsec:dft}
\subsubsection{Bulk Band Structure}

\begin{figure*}[t]
\includegraphics[width=1\textwidth]{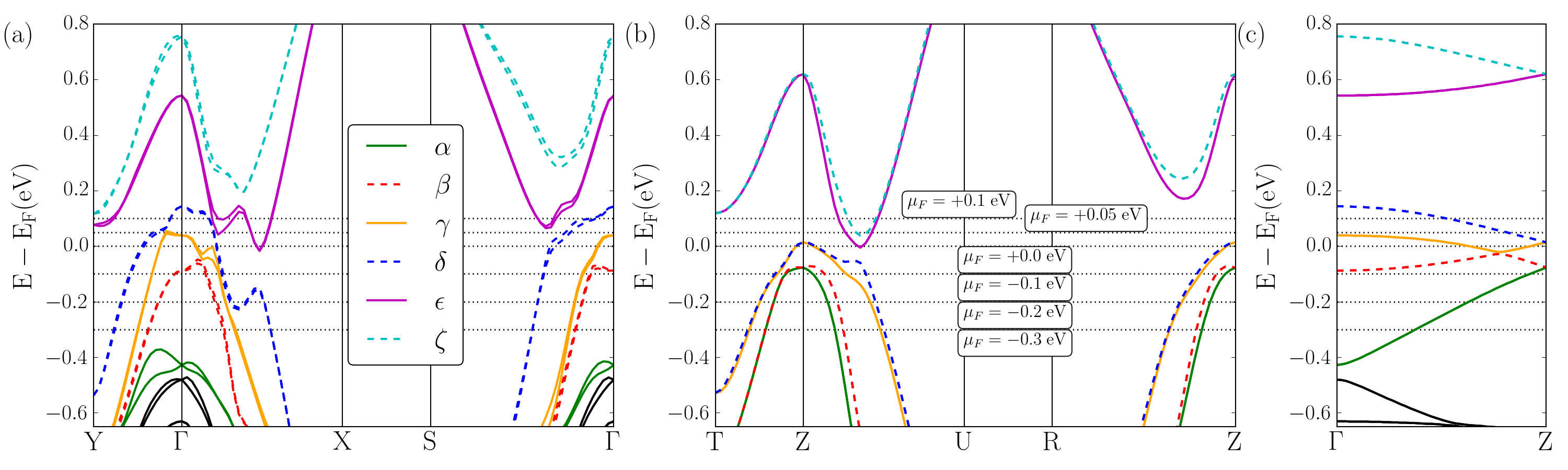} 
\caption{(a) and (b) 
Calculated band structure along high symmetry lines in the $(k_{x},k_{y})$ plane at
$k_{z}=0$  and $k_{z}=\pi/c$ , respectively. 
The 6 pairs of bands lying closest to the Fermi level are labeled and
marked [see legend in (a)].
Horizontal dashed lines 
indicate the six values for the chemical potential
considered for the Fermi surface plots in Fig.~\ref{fig:DFT-FermiS}.
(c) $k_z$ band dispersion from $\Gamma$ to Z.
}
\label{fig:DFT-bands}
\end{figure*}

First-principles calculation results for bulk T$_d$-MoTe$_2$ are now discussed to identify the expected band structure near the Fermi level. Figs. \ref{fig:DFT-bands}(a) and (b) show the band dispersion between high-symmetry points in the $(k_{x},k_{y})$ plane at $k_z=0$ and $k_z=\pi/c$, respectively, while Fig. \ref{fig:DFT-bands}(c) shows the band dispersion between the $\Gamma$ and Z points. Several bands enter an energy window between $0.3$ below and $0.1$ eV above $E_F$, within which the chemical potential $\mu_F$ could be realistically varied by doping. Specific values of $\mu_F$ that will be referred to in later sections are indicated by black dashed-lines. Six pairs of bands, labelled $\alpha$ to $\zeta$, are plotted in separate colors as per the legend shown in Fig. \ref{fig:DFT-bands}(a). Each is a Kramer's pair that forms a degeneracy at $\Gamma$. Away from TRIM points the sub-bands are separated by a k-dependent splitting induced by spin-orbit coupling (SOC). 

There are significant differences in the band structure at $k_z=0$ and $k_z=\pi/c$ due to interlayer coupling. The sub-bands are visibly split by 10-to-100 meV in most of $k_z=0$ plane, whereas they are degenerate in the $k_z=\pi/c$ plane, as expected from the symmetry considerations (see Sec.~\ref{subsec:symconstrains}). The band dispersion along $k_z$ is significant, with the bandwidth along $\Gamma$Z reaching up to $\sim 0.3$ eV for the case of the $\alpha$ bands as shown in Fig. \ref{fig:DFT-bands}(c). The $\alpha$, $\gamma$, and $\epsilon$ bands join with the $\beta$, $\delta$, and $\zeta$ bands, respectively, to form 4-fold degeneracies at the $Z$ point (see Sec.~\ref{subsec:symconstrains}).

\subsubsection{Spin-Orbital Character of Bulk Bands}

\begin{figure*}[t]
\includegraphics[width=1\textwidth]{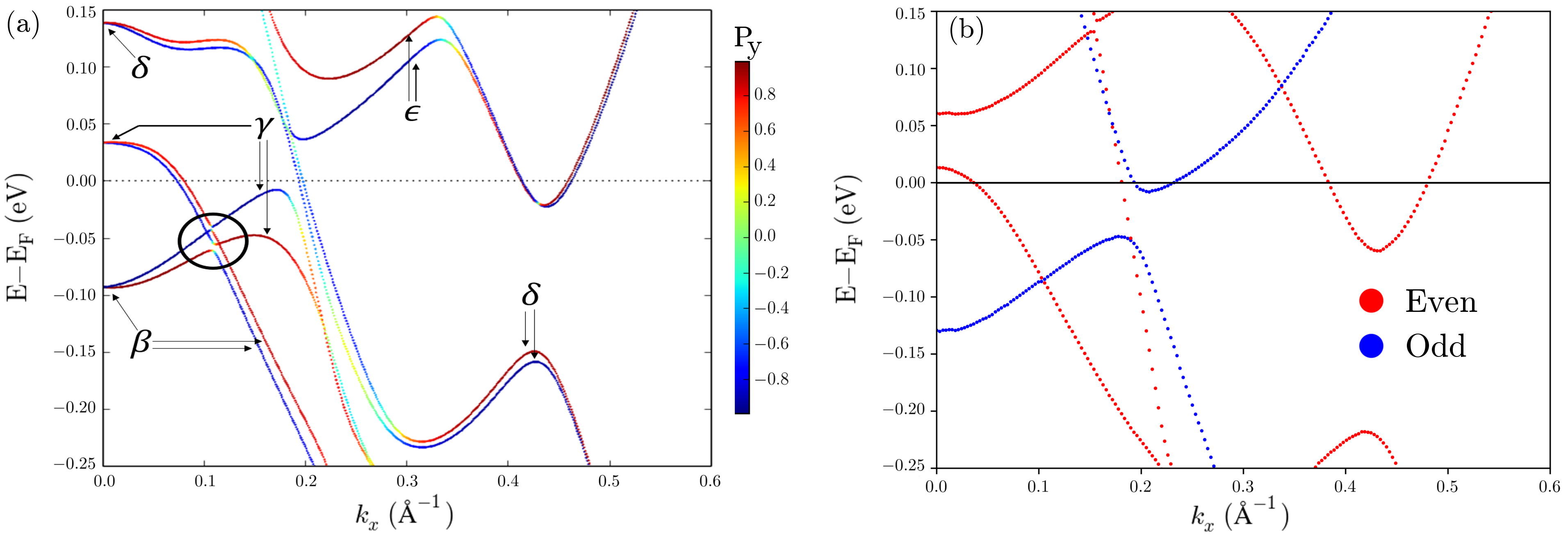} 
\caption{(a) and (b) 
Bulk band structure of T$_d$-MoTe$_2$ along the $\Gamma X$ line calculated (a) with and (b) without SOC. The color scale in panel (a) shows the calculated spin-polarization in the $y$-direction. In panel (b), the bands are plotted in red or blue to indicate the even or odd parity, respectively, of the corresponding Bloch wave function under $\widetilde{M}_y$ reflection.
}
\label{fig:bands_twopanel}
\end{figure*}

The spin and orbital texture varies between  $\alpha$ to  $\zeta$ bands. Figs. \ref{fig:bands_twopanel}(a) and (b) show the spin-resolved and non-relativistic orbital character, respectively, of the bands along the $\widetilde{M}_y$-invariant $\Gamma$X line. The \textbf{k}-dependent spin-polarization can be quantified as~\cite{azpiroz2013}
\begin{eqnarray}\label{eq:spin-pol}
\boldsymbol{P}_{n}({\bf k})=\int \Psi^{*}_{{\bf k}n}({\bf r})  
\hat{\boldsymbol{\sigma}} \Psi_{{\bf k}n}({\bf r})d^{3}r,
\end{eqnarray}
with $\hat{\boldsymbol{\sigma}}$ the Pauli spin-operator and
$\Psi_{{\bf k}n}({\bf r})$ the Kohn-Sham eigen-spinor associated with band $n$.

Fig. \ref{fig:bands_twopanel}(a) shows the spin-polarization component $P_{n}^{y}({\bf k})$ along $\Gamma \mathrm{X}$ for the bands near the Fermi level. The $x$ and $z$ components of the spin-polarization are null along this direction in the bulk BZ due to $\widetilde{M}_y$ symmetry, as discussed elsewhere \cite{Ruessmann2018}.
For the most part, the bands are 
{highly} spin-polarized along $y$ ($P_{n}^{y}({\bf k})\simeq \pm 1$), but around $k_{x}\sim 0.2$ \AA$^{-1}$ the $\delta$, $\gamma$, and $\epsilon$ states mix and their spin-polarization is reduced due to hybridization of the spin with orbitals of different symmetry.

Variation in the orbital character between the bands is also apparent in the $\widetilde{M_y}$ eigenvalues obtained from a scalar-relativistic calculation depicted in Fig. \ref{fig:bands_twopanel}(b). In this case, the orbital character is decoupled from the spin such that the $\widetilde{M}_y$ eigenvalues are even or odd as described in Sec. \ref{subsec:symm-evals}.
The detailed valence band dispersion near the Fermi level consisting of $\beta$, $\gamma$ and $\delta$ states is plotted in red-blue color code according to the even-odd $\widetilde{M}_y$ eigenvalue.
Near the $\Gamma$ point, these three pairs of bands lie close to each other and show several band-crossings such that the eigenvalues are exchanged several times. For the most part, we find two even and one odd band in this region. On the other hand, for $|\textbf{k}|>0.15$ \AA$^{-1}$\; the bands separate appreciably, and one can track that the eigenvalues for $\beta$, $\gamma$ and $\delta$ are even, odd and even, respectively.

The spin-polarization in the relativistic case is consistent with $\beta$ and $\gamma$ states possessing different orbital characters near the $\Gamma$ point ($k_x<0.1$ \AA$^{-1}$). In Fig. \ref{fig:bands_twopanel}(a) the sense of the spin-splitting between states with $+P_y$ and $-P_y$ spin-polarization is opposite between $\beta$ and $\gamma$. The intersection of these bands is circled in Fig. \ref{fig:bands_twopanel}(a). $\widetilde{M_y}$-protected crossings are found where the sub-bands have the \textit{same} spin-polarization. This can only occur when states of opposite $\widetilde{M_y}$ orbital eigenvalues cross, which is seen in the same region in Fig. \ref{fig:bands_twopanel}(b). Rapid changes in effective mass and spin-polarization are also found around the anti-crossings. Similarly, the crossing of even and odd bands above $E_F$ around $k_x=0.35$ \AA$^{-1}$ seen in  Fig. \ref{fig:bands_twopanel}(b) directly corresponds to the spin-polarization reversal found in the sub-bands of $\epsilon$ in that region of Fig. \ref{fig:bands_twopanel}(a). The considerable suppression of spin-polarization in the $\gamma$ and $\delta$ bands around $k_x=0.2-0.3$ \AA$^{-1}$, wherein $|P_y|$ is nearly zero, coincides with strong mixing of even and odd states.

\subsubsection{Predicted Chemical Potential Dependence of the Bulk Fermi Surface}

\begin{figure*}
\includegraphics[width=1\textwidth]{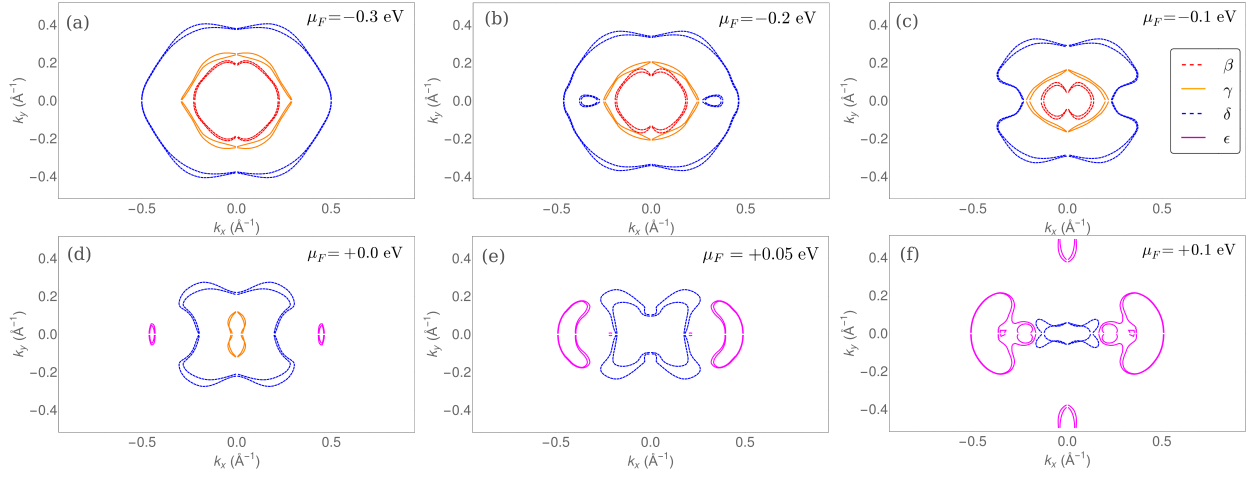} 
\caption{(a-f) Calculated Bulk Fermi surface in the ($k_{x},k_{y}$)-plane at $k_{z}=0$. Plots shown for different values of the chemical potential: 
$\mu_{F}=$ (a) $-0.3$, (b) $-0.2$, (c) $-0.1$, (d) $0.0$, (e) $+0.05$ and
(f) $+0.1$ eV. The contours are color-coded according as indicated by the legend inset within panel (c).
}
\label{fig:DFT-FermiS}
\end{figure*}

We come now to analyze the bulk Fermi surface for varying chemical potential. 
Fig. \ref{fig:DFT-FermiS} shows the Fermi surface contours in the $(k_{x},k_{y})$ plane at $k_z=0$ for different values of $\mu_F$. 
Panels \ref{fig:DFT-FermiS}(a) to \ref{fig:DFT-FermiS}(f) display the
cuts calculated for $\mu_{F}=-0.3,-0.2,-0.1,0.0,+0.05$ and $+0.1$ eV, respectively,
where the effective chemical potential shift is relative to the
calculated Fermi level ($\mu_{F}=0$). 
Only four pairs of bands present contours in this regime, namely $\beta,\gamma,\delta$ and $\epsilon$. 
Let us proceed to discuss from lowest to highest values of the chemical potential. For negative values [Figs. \ref{fig:DFT-FermiS}(a)-(c)], three hole pockets originating from the $\beta,\gamma$ and $\delta$ bands are centered around the $\Gamma$ point. All grow in size as the chemical potential is lowered.
The outermost contours correspond to $\delta$ and are nearly circular. 
The $\beta$ and $\gamma$ contours form the inner contours and are roughly elliptical, elongating near the $k_y=0$ line.
An exception exists at $\mu_{F}= -0.2$ eV [Fig. \ref{fig:DFT-FermiS}(b)] where two small hole pockets appear between the 
$\gamma$ and outer $\delta$ contours. 
These pockets originate from
the ``tilted S-shape'' in the band dispersion of the $\delta$ states [see Fig. \ref{fig:DFT-bands}(a)].
At $\mu_F=0$ [Fig. \ref{fig:DFT-FermiS}(d)], the $\beta$ states become absent and the $\delta$ states produce highly anisotropic,``palmier-shaped'' \cite{Belopolski2016} contours with a large spin-splitting reaching up to $\approx$0.1 \AA$^{-1}$ near the $\Gamma$S direction, while $\epsilon$ states give rise to small, elliptical electron pockets midway between $\Gamma$ and X. 
As $\mu_F$ is raised to +0.05 eV [Fig. \ref{fig:DFT-FermiS}(e)], the contours are exclusively produced by the $\delta$ and $\epsilon$ bands. The $\delta$ contours contract within the $k_y$ axis and their spin-splitting along the $\Gamma-S$ direction increases while the electron pockets formed by the $\epsilon$ states opens to form ``kidney-shaped'' contours. An additional pair of very small electron pockets, also derived from the $\epsilon$ states, emerge just outside of the $\delta$ contours on the $k_x$-axis. At $\mu_F=+0.1$ eV [Fig. \ref{fig:DFT-FermiS}(f)], there is a strong spin-dependence of the contour shapes between $k_x = 0.1$ and $0.3$ \AA$^{-1}$ and 12 distinct pockets produced by the $\epsilon$ band appear. The $\delta$ contours contract toward the $k_y=0$ axis, with an inner contour forming an elliptical shape and the outer contour forming ``bow tie'' shape.
As a general conclusion, Fig.~\ref{fig:DFT-FermiS} predicts a roughly isotropic Fermi surface for negative values of the doping,
while for positive values it becomes anisotropic with an elongated shape around the $k_{y}=0$ axis.

\subsubsection{Electronic Structure of the (001) Surface}
\label{subsec:dft-surface}

\begin{figure}
\includegraphics[width=0.4\textwidth]{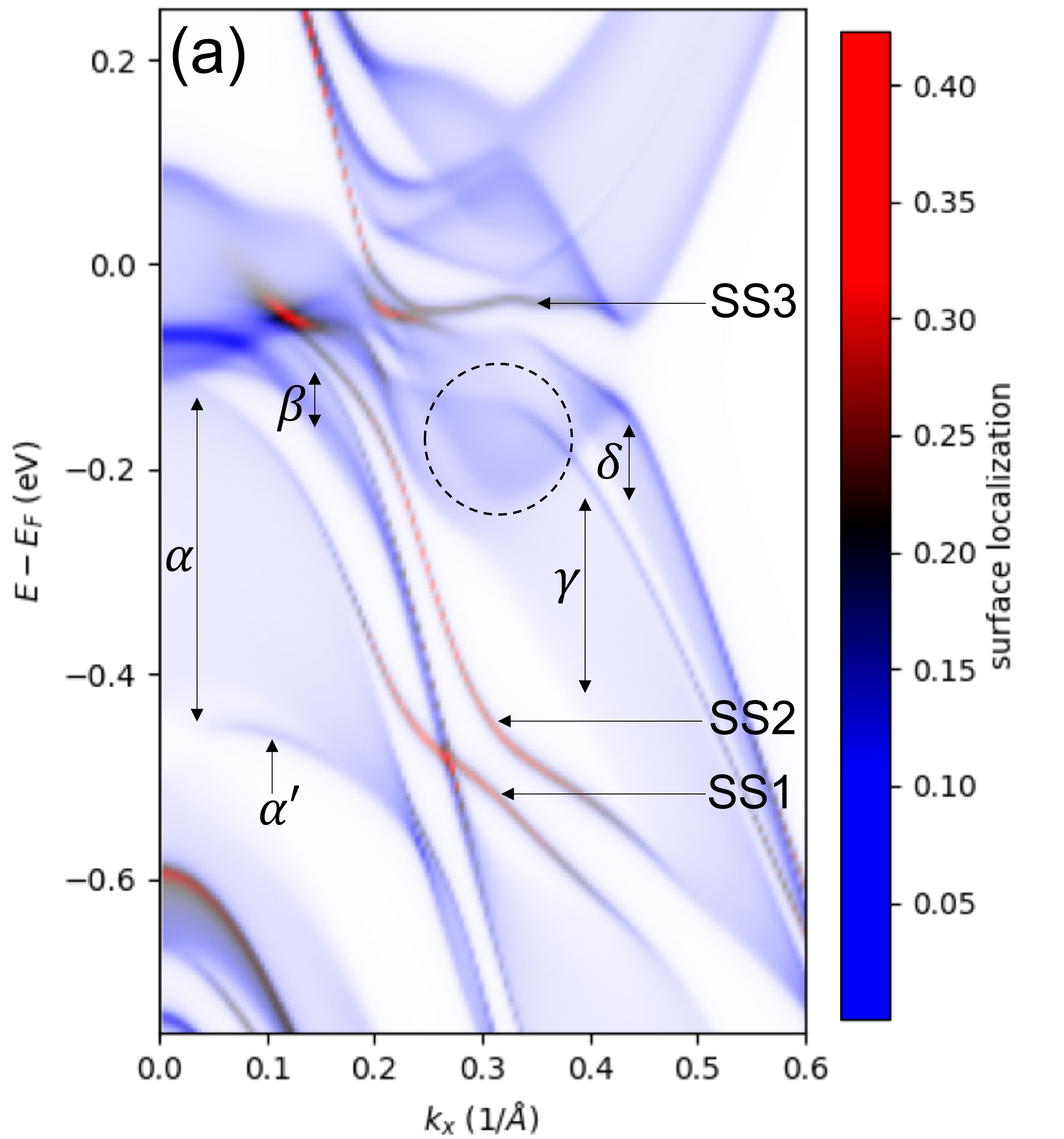} \includegraphics[width=0.58\textwidth]{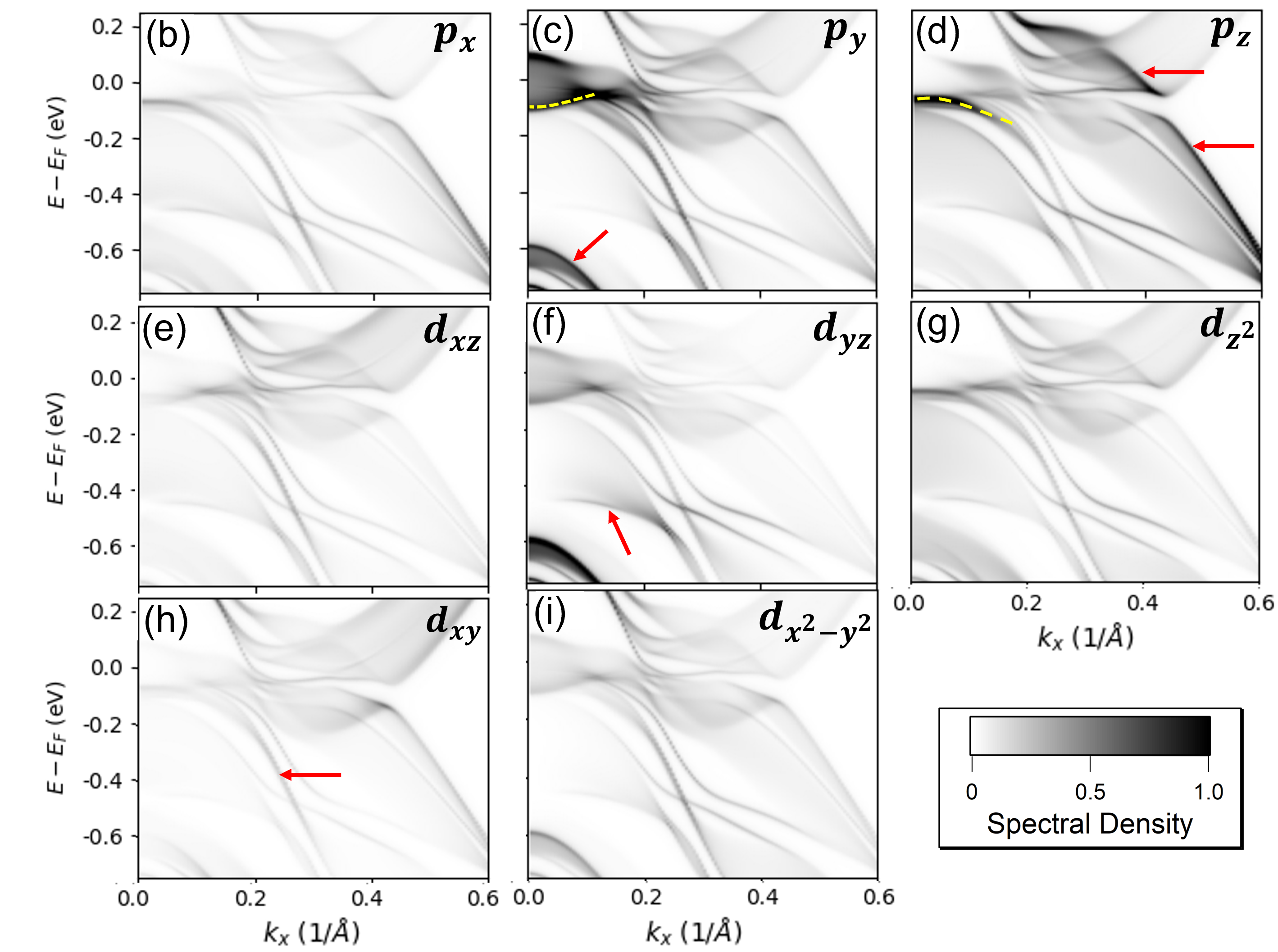}
\caption{Dispersion of spectral density along $\overline{\Gamma \mathrm{X}}$ calculated for semi-infinite T$_d$-MoTe$_2$(001). 
(a) Surface localization of the spectral density. The color scale (inset) indicates the degree of surface localization and the color opacity is proportional to the total density. 
(b-i) Orbital-projected spectral densities in grayscale. The scale is shown in the inset, where 1 corresponds to the maximum orbital-projected density for the given orbital in each plot. }
\label{fig:DFT-surface-bands}
\end{figure}

As the last step in the theoretical analysis, we now focus on describing  the electronic structure of the (001) surface of MoTe$_2$, which will facilitate interpreting the surface resonances 
observed experimentally.
Fig. \ref{fig:DFT-surface-bands}(a) shows the dispersion of the spectral density within the MoTe$_2$ trilayers along the $\overline{\Gamma \mathrm{X}}$ direction. High and low densities correspond to opaque and translucent coloring, respectively. The blue-red color scale (inset) indicates the degree of localization in the surface MoTe$_2$ layer. Double-arrows indicate the bandwidth of each of the bulk states that are projected into the SBZ. Consistent with $\Gamma$Z dispersion shown in Fig. \ref{fig:DFT-bands}(c), the $k_z$ dispersion of the $\alpha$ states produces a continuum of spectral density from just below 0.4 eV up to 0.1 eV around $\overline{\Gamma}$. The $\gamma$ states also have significant bandwidth of up to 0.2 eV near $k_x=0.45$ \AA$^{-1}$. Referring back to Fig. \ref{fig:DFT-bands}, the local energy minimum in the ``tilted S-shape'' of the $\beta$ states lies below the energy maximum of the $\gamma$ states at $k_z=\pi/c$. We therefore see that the projections of $\beta$ and $\gamma$ overlap around $k_x=0.4$ \AA$^{-1}$. Some enhancement of the spectral density is seen within that region, which is circled with a black dashed-line in Fig. \ref{fig:DFT-surface-bands}(a). Outside of this region, the bandwidth of the $\delta$ states narrows significantly. The $\beta$ states produce an especially narrow bandwidth of only a few meV in the range of $k_x=0.25-0.3$ \AA$^{-1}$. Three surface states clearly separate from the segments of the bulk continuum labelled SS1, SS2, and SS3. Each is indicated by arrows in Fig. \ref{fig:DFT-surface-bands}. SS3 produces a topologically trivial Fermi arc \cite{Tamai2016,Crepaldi2017,Weber2018}. Hole-like surface states SS1 and SS2 appear farther below $E_F$ and have also been studied in detail previously \citep{Crepaldi2017}.

The corresponding projection of spectral density onto Te $5p$ and Mo $4d$ orbitals is shown in Fig. \ref{fig:DFT-surface-bands}(b-i). The distribution of spectral weight varies from image to image, but we note that the weight due to $\beta$ remains visible in all of the plots in the 0.2-0.4 eV below $E_F$, as indicated by red arrow in Fig. \ref{fig:DFT-surface-bands}. The density contributed by the $\alpha$ states near $k_x=0$ varies significantly and is most pronounced for the $p_x$, $d_{xz}$, $p_z$, and $d_{z^2}$ projections. Surface spectral density on the lower edge of the $\alpha$ continuum is labelled $\alpha'$ in Fig. \ref{fig:DFT-surface-bands}(a). The region around this edge has relatively large $d_{yz}$-orbital character indicated by an arrow in Fig. \ref{fig:DFT-surface-bands}(f). The $\gamma$ and $\delta$ continua are fully visualized only in the $p_z$-orbital-projected density shown in Fig. \ref{fig:DFT-surface-bands}(d). The $p_z$ character is especially enhanced in the upper edges of $\gamma$ and $\delta$ and the inner part of the electron pocket formed by $\epsilon$, which is indicated by arrows in Fig. \ref{fig:DFT-surface-bands}(d). The dispersion of spectral density close to $E_F$ at $k_x=0$ can also appear to be electron-like or hole-like, depending on the orbital projection, as indicated by yellow dashed-lines in Fig. \ref{fig:DFT-surface-bands}(c) and (d), respectively. Finally, the arrow in Fig. \ref{fig:DFT-surface-bands}(c) indicates hole-like states lying just below $\alpha$ with a band maximum around 0.6 eV below $E_F$ that are overwhelmingly of $p_y$ and $d_{yz}$ character. From these results and considering the ARPES selection rules discussed in Ref. \cite{Ono2021}, the ARPES spectra are expected to change significantly with the light-polarization and experimental geometry.

\subsection{Experimental Results}
\label{subsec:exp}

\subsubsection{ARPES of T$_d$-MoTe$_2$(001) with Variable Photon Energy and Polarization}

\begin{figure*}
\includegraphics[width=1\textwidth]{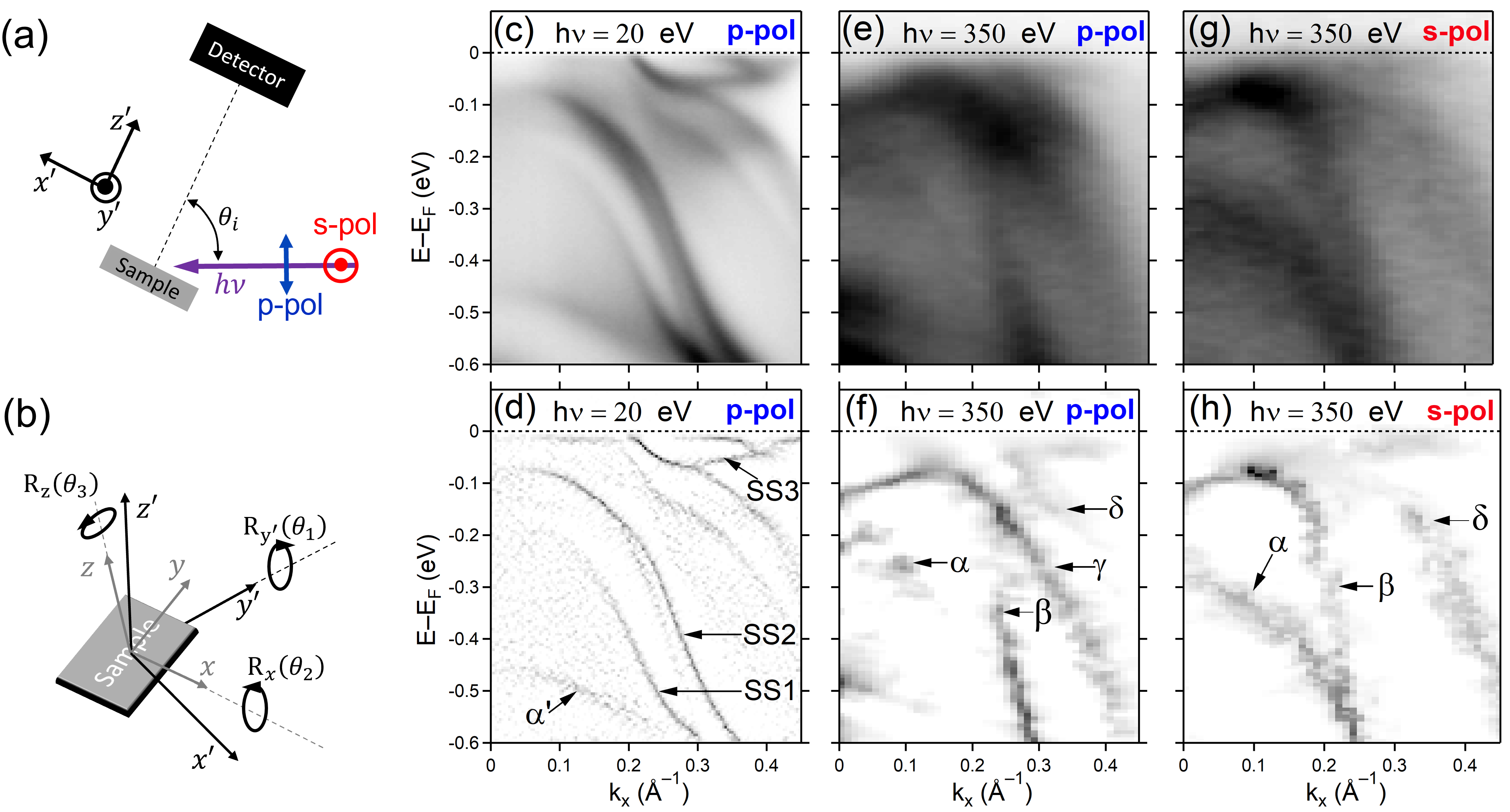} 
\caption{(a-b) Schematics of the photoemission experiment geometry. (c-d) UV-ARPES and (e-h) SX-ARPES data collected from the $x=0$ sample using (c-f) $p$- and (g-h) $s$-polarized photons. ARPES intensity along $\overline{\Gamma \mathrm{X}}$ is shown in (c), (e), and (f) above the corresponding 2D curvature in (d), (f), and (h), respectively.
}
\label{fig:ARPES-bands}
\end{figure*}

We now proceed to discuss the experimental ARPES results. 
We achieved full characterization of the bulk and surface bands by changing the photon energy and light-polarization. We used either $p$- or $s$- polarization, which is parallel or perpendicular to the scattering plane of the light, respectively, as illustrated in Fig. \ref{fig:ARPES-bands}(a). A fixed laboratory reference frame is defined in terms of primed position coordinates $(x',y',z')$ and corresponds to the sample orientation when the surface normal is aligned to the center of the detector entrance along the $z'$-axis. The sample reference frame is defined by the $(x,y,z)$ axes discussed previously in Section A. The sample is manipulated by the sequence of rotations $R_{y'}(\theta_1)$, $R_x(\theta_2)$, and $R_z(\theta_3)$ about the $y'$, $x$, and $z$ axes by angles $\theta_1$, $\theta_2$, and $\theta_3$, respectively, as illustrated in Fig. \ref{fig:ARPES-bands}(b). The light approaches the sample surface in the $x'z'$-plane at an angle $\theta_i$ from the $z'$-axis, which was 45\textdegree~and 70\textdegree~for the UV- and SX-ARPES experiments, respectively, such that the vector potential has $x'$ and $z'$ components for $p$-polarization, whereas only the $y'$ component is finite for $s$-polarization. $\theta_1$ and $\theta_2$ were always $<5$\textdegree. Measurements were taken with either $\theta_3=0$ or $\theta_3=90^{\circ}$ within a few-degree accuracy. The figure captions explicitly state cases where  $\theta_3=90^{\circ}$ was used.

Writing the vector potential of the incident beam as
$\Avec = A_0\,\epsh\,e^{i(\vec k^{ph}\cdot\vec r-\omega t)}$, where $k^{ph}$ is the photon wave-vector, the polarizations
described above correspond to $\epsh_s=\hat y'$ and
$\epsh_p=-\cos(\theta_i)\hat x'+\sin(\theta_i)\hat z'$ in the laboratory frame. Expressed in
the sample reference frame, and setting $\theta_1=\theta_2=0$, these become
\begin{align*}
\epsh_s&=\sin(\theta_3)\,\hat x+\cos(\theta_3)\,\hat y,\\
\epsh_p&=-\cos(\theta_3)\cos(\theta_i)\,\hat x+\sin(\theta_3)\cos(\theta_i)\,\hat y
+\sin(\theta_i)\,\hat z .
\end{align*}
Most importantly, s-polarized light has no component along the sample normal for any $\theta_3$, whereas p-polarized light retains one of magnitude
$\sin(\theta_i)$. It is this presence or absence of $A_z$, which switches the excitation
from $p_z$-derived initial states on or off (as described in Appendix~\ref{sec:App-A}), that is principally exploited in the
polarization-dependent measurements for the purpose of disentangling states of different orbital character.

Circular light-polarization (c-pol) was additionally used for the photon-energy-dependent measurements discussed below. The two circular polarizations are the combinations
\begin{equation*}
\epsh_\pm=\tfrac{1}{\sqrt2}\bigl(\epsh_p\pm i\,\epsh_s\bigr),
\end{equation*}
so that right- and left-handed light differ only in the sign of the $90^\circ$ phase shift
between the components of $\Avec$ lying within and perpendicular to the scattering plane.
This sign sets the sense in which $\Avec$ rotates about the propagation direction: taking the real part, $\Avec(t)\propto\epsh_p\cos(\omega t)\pm\epsh_s\sin(\omega t)$, so that $\epsh_+$
($\epsh_-$) corresponds to a right-handed (left-handed) rotation about $\hat k^{ph}$. We denote these $C^+$ and $C^-$. Each contains $s$ and $p$ components with equal magnitude. We employed c-pol to mitigate linear dichroism when desired, rather than study any chiral response of the material (circular dichroism), which is outside the scope of this work. Both $C^+$ and $C^-$ were used in the measurements reported here, as indicated in the text and figure captions.

\begin{figure*}
\includegraphics[width=0.85\textwidth]{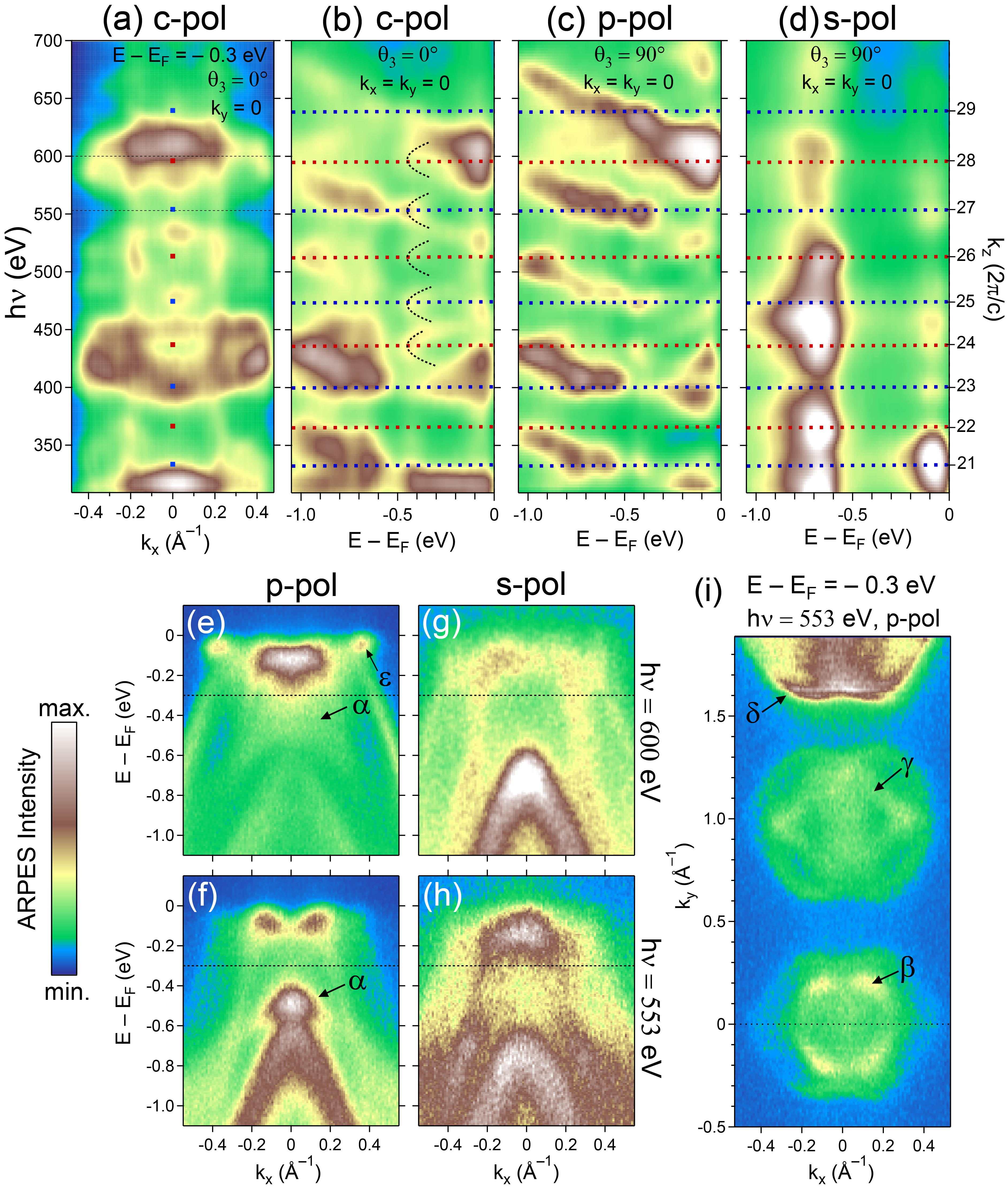}
\caption{Photon energy dependent SX-ARPES data for the $x=0$ sample. (a) Distribution of ARPES intensity along the $k_x$-axis at 0.3 eV binding energy collected using $C^+$-polarized light. (b-d) Energy distribution of ARPES intensity at $\overline{\Gamma}$ collected using (b) $C^+$, (c) $p$-, and (d) $s$-polarized light. The red (blue) markers in (a-d) indicate centers of even (odd) numbered Brillouin zones. (e-h) Band structure images near two different $\Gamma$-points collected with (e,g) 600 eV and (f,h) 553 eV photons with (e,f) $p$- and (g,h) $s$-polarization. (i) Constant energy contours of $\beta$, $\gamma$, and $\delta$ hole states imaged in successive surface Brillouin zones collected using 553 eV, $p$-polarized light. The data in panels (a-b) and (c-h) were collected in the $\theta_3=0$ and $\theta_3=90^{\circ}$ geometry, respectively.}
\label{fig:ARPES-hn}
\end{figure*}

Fig. \ref{fig:ARPES-bands}(c-h) and Fig. \ref{fig:ARPES-hn} show ARPES spectra along $\overline{\Gamma \mathrm{X}}$ collected from the $x=0$ sample at temperatures below 30 K, where the crystal is in the T$_d$ structural phase. The high-resolution UV-ARPES spectrum in Fig. \ref{fig:ARPES-bands}(c) was collected using $p$-polarized 20 eV photons. The photoemission intensity therein closely resembles the calculated surface spectral density in Fig. \ref{fig:DFT-surface-bands}(a), with sharp surface bands appearing over diffuse bulk states. The corresponding 2D curvature is plotted below in Fig. \ref{fig:ARPES-bands}(d) to highlight the peak intensity of dispersive features \cite{Zhang_2D_curv}. The surface bands SS1, SS2, and SS3 [see Sec. \ref{subsec:dft-surface}]
are easily identified and are indicated by arrows. The surface resonance at the bottom edge of the continuum of $\alpha$ states is additionally indicated and surface spectral density at the upper edges of the $\gamma$ and $\delta$ states is also visible in the curvature image. In Fig. \ref{fig:ARPES-bands}(c), a broad ``tilted S'' pattern of intensity also appears in the region 0.1-0.2 eV below $E_F$ between $k_x=0.25$ and $0.45$ \AA$^{-1}$, in qualitative agreement with DFT results shown in Fig. \ref{fig:DFT-surface-bands}(a). This corresponds to the surface resonances lying in the overlap of the $\gamma$ and $\delta$ band dispersion discussed before.

The bulk-derived states are prominent in the SX-ARPES data shown in Fig. \ref{fig:ARPES-bands}(e-h). SX-ARPES is generally more bulk-sensitive than UV-ARPES due to the increased mean free path of the photoelectrons \cite{Strocov2003}. Although surface states can be resolved by SX-ARPES \cite{Queiroz2017}, here they are not distinguishable. This could be due to relatively weak photoemission intensity and/or because their closeness to bulk states combined with the energy resolution of this experiment. There is ARPES intensity near $E_F$ at $k_x>0.3$ \AA$^{-1}$ that is likely contributed by both the $\epsilon$ states and SS3. All four of the hole-like bands are observed in the ARPES spectra produced using $p$-polarized light shown in Fig. \ref{fig:ARPES-bands}(e). The $\alpha$, $\beta$, $\gamma$, and $\delta$  states are each indicated by arrows in the corresponding 2D curvature plot in Fig. \ref{fig:ARPES-bands}(f). The $\alpha$ states only show diffuse intensity around 0.2 eV below $E_F$, which is near the upper edge of the $\alpha$ bandwidth. Upon changing to $s$-polarization, the intensity derived from the $\alpha$ band shifts to form a clear hole-like dispersion with a maximum around 0.3~eV below $E_F$. The intensity contributed by the $\beta$ states remains strong, but the $\gamma$ states effectively disappear, while the $\delta$ states are better resolved at higher binding energies. 
{The apparent polarization-dependent displacement of the $\alpha$ band may arise from the overlap and hybridization of the $\alpha$ and $\beta$ states, combined with polarization-dependent matrix elements and finite $k_z$ resolution.} 

Further variation of the photon energy and polarization helps to understand these effects. Fig. \ref{fig:ARPES-hn}(a) and (b) show the momentum distribution of photoemission intensity along $\overline{\Gamma \mathrm{X}}$ at 0.3~eV below $E_F$ and the energy distribution of the ARPES intensity at the $\overline{\Gamma}$ point, respectively, collected as a function of photon energy. These scans were collected with $C^+$ light, to mitigate the linear dichroism effect noted earlier. Markers in Fig. \ref{fig:ARPES-hn}(a-d) indicate points at which the wave-vector of the final state is a multiple of a reciprocal lattice vector $2\pi n_z/c$. These correspond to the centers of successive BBZs $\Gamma_{0,0,n_z}$ and are plotted in {red} and {blue} for even and odd $n_z$, respectively. The dispersion of the $\alpha$ states between $0.45$~eV and $0.2$~eV binding energy is visible in most of the BZs in Fig. \ref{fig:ARPES-hn}(b). Black dashed lines are drawn over several periods of the $\alpha$ dispersion minima as a guide to the eye. The $\alpha$ states contribute splotches of intensity around $k_x=0$ in the constant energy map Fig. \ref{fig:ARPES-hn}(a), wherein the photoemission intensity of $\beta$ states weave slightly around $+0.2$ and $-0.2$ \AA$^{-1}$, and $\gamma$ and $\delta$ states appear outside of that. 

To examine the light-polarization dependence further, the data shown in Fig. \ref{fig:ARPES-hn}(c-h) were collected using a different experimental geometry by rotating the sample in-plane by $\theta_3=90$\textdegree. In this case, the vector potential is along the sample $x$-axis for $s$-polarized light and has $\hat{y}$ and $\hat{z}$ components for $p$-polarized light. 
Some of the observed features are strongly affected by light polarization. The energy dispersion of the $\alpha$ states at $\overline{\Gamma}$ can still be seen in Fig. \ref{fig:ARPES-hn}(c) when $p$-polarization is used, but these states effectively disappear from the spectra across the full photon energy range in Fig. \ref{fig:ARPES-hn}(d) when $s$-polarization is used.

For the purpose of compactly showing an experimental identification of the $\beta$, $\gamma$, and $\delta$ constant energy contour shapes below $E_F$ that were previously described in Fig.~\ref{fig:DFT-FermiS}(a-b), Fig.~\ref{fig:ARPES-hn} displays a map at $E-E_F=-0.3$ eV containing three SBZs collected with $p$-polarized 553 eV photons. In the first SBZ (around $k_x=k_y=0$), the circular, inner $\delta$ contour appears prominently, whereas the triangular $\gamma$ contour and outer, elliptical $\delta$ contour are pronounced in the second and third SBZs along the $k_y$-direction, respectively. This provides another demonstration that the relative intensities of different bands can be strongly modulated by the length of the photoelectron wave-vector, and shows that the general shapes and sizes expected from contours that will make up the Fermi surface of the Nb-doped sample are in qualitative agreement with the theory discussed previously.

\subsubsection{Experimental Comparison of $x=0$ and $x=0.18$ Electronic Structure}

\begin{figure*}
\includegraphics[width=1\textwidth]{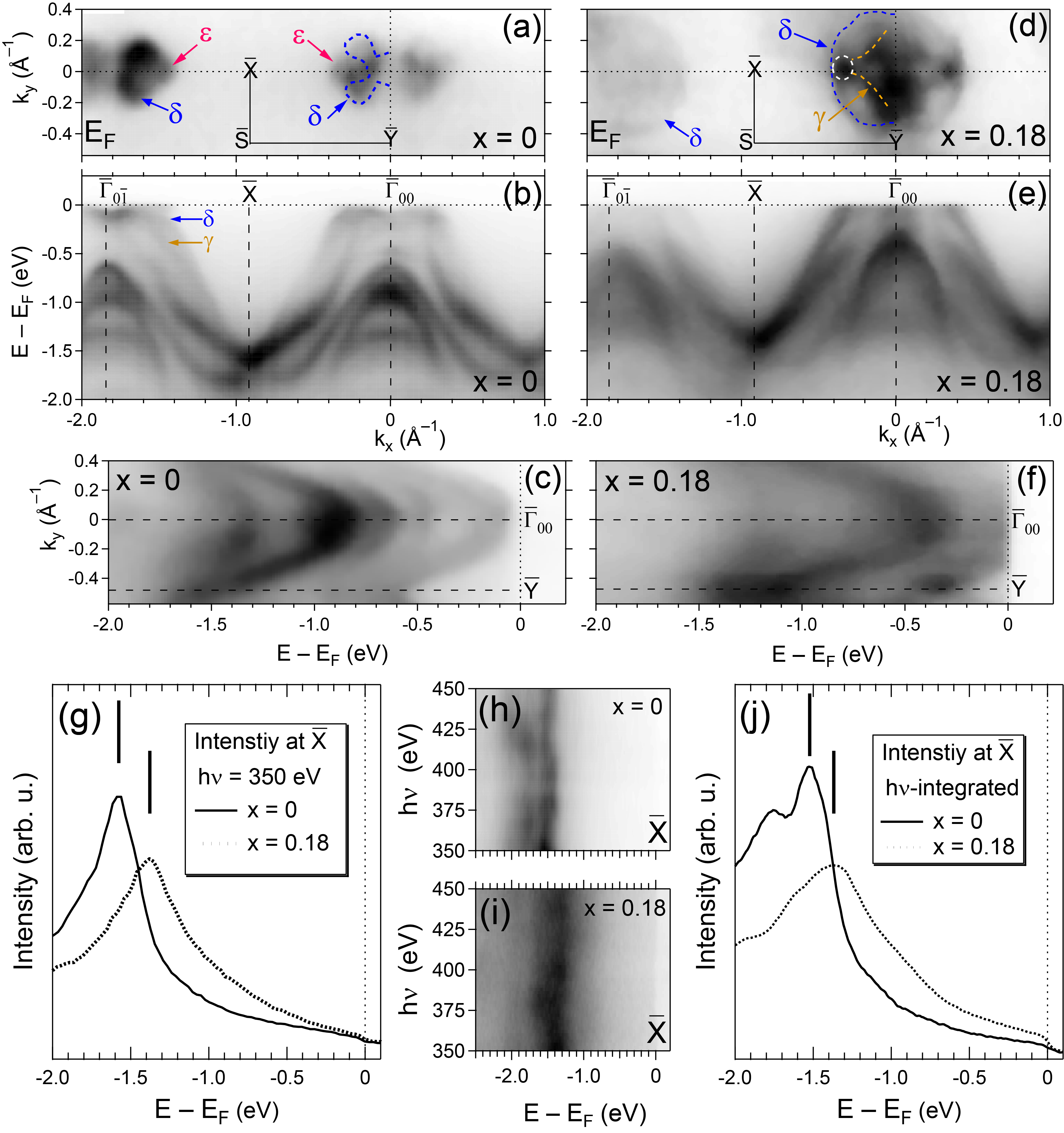}
\caption{{(a-g) SX-ARPES spectra of (a-c) $x = 0$ and (d-f) $x = 0.18$ samples kept at $T\approx15$ K collected using $p$-polarized 350 eV photons. 
ARPES intensity maps of the (a,d) Fermi surface and band structure along the (b,e) $k_{y}=0$ and (c,f) $k_{x}=0$ line. 
(g) Binding energy distributions of the ARPES intensity at the $\overline{\mathrm{X}}$ point. 
(h-i) Photon-energy-dependent ARPES intensity at $\overline{\mathrm{X}}$ for (a-c) $x = 0$ and (d-f) $x = 0.18$ samples.
(j) Distributions of the photon-energy-integrated intensity from 350-450 eV.  
Solid and dotted-lines in (g,j) correspond to $x = 0$ and $0.18$ cases, respectively.
Panels (e) and (f) show that the pockets 
for $x=0.18$ are formed by hole-like bands.
}}
\label{fig:ARPES-overview}
\end{figure*}

There are significant differences in the SX-ARPES data collected from $x=0$ and $0.18$ samples, which are displayed in Fig. \ref{fig:ARPES-overview}(a-c) and Fig. \ref{fig:ARPES-overview}(e-f), respectively. The measurements cover a large area of momentum space containing several SBZs. The boundary of the (001) SBZ is indicated by solid black lines in the Fermi surface maps in Fig. $\ref{fig:ARPES-overview}$(a) and Fig. $\ref{fig:ARPES-overview}$(d). High-symmetry points $\overline{X}$, $\overline{Y}$, $\overline{S}$ on the boundary are labeled, along with the center of the first SBZ $\Gamma_{00}$ and second SBZ $\Gamma_{0\overline{1}}$. The difference in the sizes of Fermi pockets between the $x=0$ and $0.18$ case is very clear. The outer boundaries of the pockets for the $x=0$ sample are indicated by dashed-lines in Fig. \ref{fig:ARPES-overview}(a) and correspond to $\delta$ and $\epsilon$ states. Their size and shape compare well with the DFT Fermi contours in $\mu_F=0.05$ eV in Fig. \ref{fig:DFT-FermiS}(e). The $x=0.18$ sample has a much larger Fermi surface where the outermost contour, formed by the $\delta$ states, encloses $\overline{\Gamma}$ and is outlined by blue dashed-lines in Fig. \ref{fig:ARPES-overview}(d). It is roughly elliptical in shape with an average radius of about 0.4 \AA$^{-1}$. This corresponds to $\mu_F=-0.2$~eV in Fig. \ref{fig:DFT-FermiS}(b). 

Other pockets appear within this contour, including a triangular pocket arising from the $\gamma$ states, indicated by yellow dashed-lines. An enhancement of intensity observed at the edge of this feature is circled by a white dashed-line and corresponds to the resonance of the $\gamma$ and $\delta$ states that was described by the surface calculations. Some intensity contributed by the $\alpha$ states appears close to $\overline{\Gamma}$. ARPES spectra along $\overline{\Gamma X}$ and $\overline{\Gamma Y}$ directions of the SBZ in Fig. \ref{fig:ARPES-overview}(e) and Fig. \ref{fig:ARPES-overview}(f) show that all of the Fermi pockets of the $x=0.18$ sample are formed by hole-like bands. An upward shift of all of the bands is apparent from comparison with the corresponding data from the $x=0$ sample shown in Fig. \ref{fig:ARPES-overview}(b) and (c). 

To provide an estimate of the energy shift, the energy distribution curves (EDCs) of the ARPES intensity at $\overline{X}$ are plotted in Fig. \ref{fig:ARPES-overview}(g). The EDCs are plotted as solid (dotted) lines for the $x=0$ ($x=0.18$) case. The difference in the valence band energy maxima (marked by vertical bars) for the two different samples is $0.20$ eV. The peak intensity at $\overline{X}$ undergoes slight dispersion with photon energy ($k_z$) in both samples, as seen in Fig. \ref{fig:ARPES-overview}(h-i). The band shift is therefore appreciable across the 350-450 eV photon energy range, and persists in the photon-energy-integrated EDCs shown in Fig. \ref{fig:ARPES-overview}(j). As would be expected for an alloy with substitutional disorder, the $x=0.18$ photon-integrated spectra show broader features and a more-significant background corresponding to decreased quasiparticle lifetimes and inelastic photoelectron scattering, respectively. The shift in the maximum intensity of the photon-integrated spectra compared between the two cases is $0.15$ eV. The effect of strong inelastic background in the $x=0.18$ case likely contributes to a reduction in the peak-to-peak shift. Nevertheless, we can conclude that the {quasi-rigid} band shift is $\approx0.2$ eV, within the SX-ARPES experiment resolution ($<0.1$ eV). Clean observation of the band shift requires isolating the signal of a given band and measuring its binding energy at equivalent 3D momentum coordinates for both samples, which is done at the end of the next section.    

\subsubsection{ARPES for $x=0.18$ with Variable Photon Energy and Polarization}
\begin{figure*}
\includegraphics[width=0.9\textwidth]{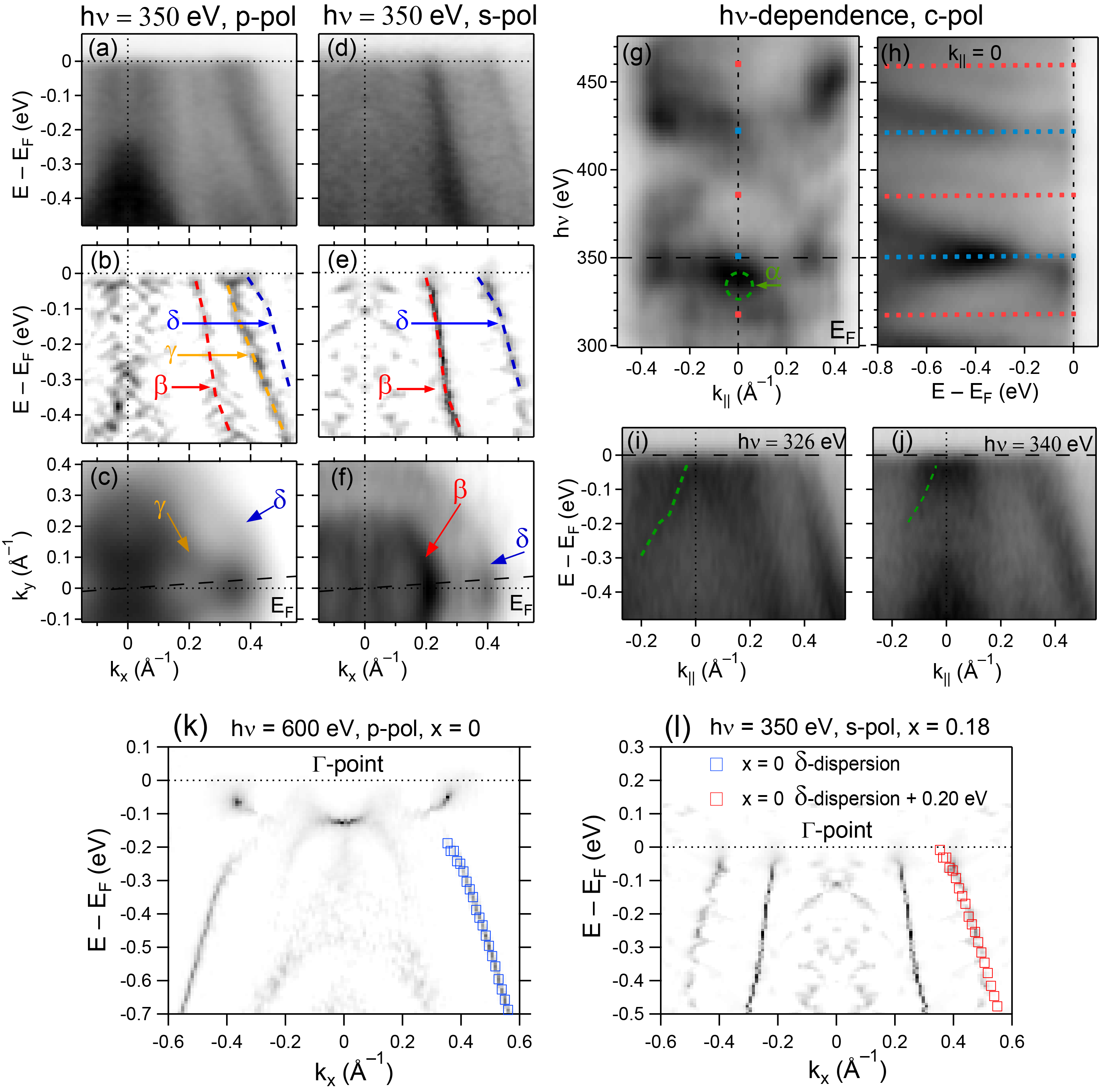} 
\caption{(a-j) SX-ARPES data for the $x=0.18$ sample. (a-f)  Data collected 350 eV photons with (a-c) $p$- and (d-f) $s$-polarization. (a,d) ARPES intensity along the $\overline{\Gamma X}$ line and (b,e) its 2D curvature. (c,f) Constant energy cuts of the ARPES intensity at $E_F$. (g-j) Photon energy dependence of the ARPES intensity collected with $C^-$ light polarization. (g) Momentum distribution of intensity at $E_F$. (h) Energy distribution of intensity at $\overline{\Gamma}$. The light-red (light-blue) markers in (g-h) indicate centers of even (odd) numbered Brillouin zones. (i-j) Band dispersion maps along $k_{||}$ collected using (i) 326 and (j) 340 eV. The $k_{||}$ axis is indicated by long dashed lines in panels (c) and (f). Green, red, yellow and blue arrows and dashed-lines guide the eye to spectral features from the $\alpha$, $\beta$, $\gamma$, and $\delta$ states, respectively. (k-l) Side-by-side comparison of 2D curvature of the ARPES intensity for (k) $x=0$ and (l) $x=0.18$ for conditions resolving the $\delta$ states measured along $k_x$, where the selected photon energies place both measurements near the $\Gamma$-point.
}
\label{fig:ARPES-Nb}
\end{figure*}

The detailed low-energy electronic structure of the $x=0.18$ sample is examined in Fig. \ref{fig:ARPES-Nb}(a-j). First, we will identify the dispersion of the $\beta$, $\gamma$, and $\delta$ states and show that they exhibit linear dichroism in the same way as for the $x=0$ sample discussed previously. Fig. \ref{fig:ARPES-Nb}(a) and Fig. \ref{fig:ARPES-Nb}(b) show the measured band dispersion along $\overline{\Gamma X}$ observed by SX-ARPES using $p$- and $s$-polarized 350 eV photons, respectively. The corresponding 2D curvature plots are shown in the panels directly below [Fig. \ref{fig:ARPES-Nb}(b,e)] with the dispersion of the $\beta$, $\gamma$, and $\delta$ states indicated by color-coded arrows and dashed-lines. The dispersion of all of these states is visible with $p$-polarization. The $\gamma$ and $\beta$ states present dominant and relatively faint ARPES intensity, respectively. With $s$-polarization, the $\gamma$ states effectively disappear while the intensity of the $\beta$ states becomes substantial. This effect was previously observed in the ARPES data of the $x=0$ sample that was shown in Fig. \ref{fig:ARPES-bands} and is apparent in the Fermi surface plots shown in Fig. \ref{fig:ARPES-Nb}(c,f). The triangular contour of the $\gamma$ states is clearly observed with $p$-polarized light. With $s$-polarization, the triangular contour disappears whereas the rounded contour of the $\beta$ states is clear. Meanwhile, the outermost contour in the Fermi surface, belonging to the $\delta$ states, is present with a faint ARPES intensity under both polarizations.

Photon energy dependent SX-ARPES data collected using $C^-$-polarized light are presented in Fig. \ref{fig:ARPES-Nb}(g-f). The ARPES intensity of a slice of the Fermi surface along the $k_{||}$ axis is mapped as a function of photon energy in Fig. \ref{fig:ARPES-Nb}(g), where $k_{||}$ axis is slightly rotated in the $k_x,k_y$ plane by 4\textdegree with respect to the $k_x$-axis and is drawn over the previous Fermi surface plots in a long-dashed line. The momentum-space location and photon energy dispersion of these states is similar to the results shown for the $x=0$ sample at 0.3 eV binding energy in Fig. \ref{fig:ARPES-hn}(a). The $\alpha$ states contribute intensity around $k_{||}=0$. The $\beta$ and $\delta$ states remain close to $k_{||}=0.2$ and $0.4$ \AA$^{-1}$, respectively, while the intensity contributed by the $\gamma$ states weaves in the space between. Points at which the wave-vector of the photoemission final state is a multiple of a reciprocal lattice vector $2\pi n_z/c$ are indicated by light-red and light-blue markers for even and odd $n_z$, respectively. The same points correspond to the $\Gamma$ point within successive BBZs and are represented by color-coded markers in the energy dispersion of the ARPES intensity at $k_{||}=0$ that is shown in Fig. \ref{fig:ARPES-Nb}(h). As discussed in previous sections, the energy maxima and minima of $\alpha$ states occur at the $\Gamma$ and $\mathrm{Z}$ points of the BBZ, respectively. This creates Fermi surface pockets enclosing the $\mathrm{Z}$ point. The Fermi pocket expected for the $\alpha$ states is sketched by a green dashed line in Fig. \ref{fig:ARPES-Nb}(g). The energy dispersion of ARPES intensity along $k_{||}$ collected at selected photon energies in Fig. \ref{fig:ARPES-Nb}(i-j). 

Looking back to the data in Fig. \ref{fig:ARPES-Nb}(a) collected with 350 eV photons, some intensity from the $\alpha$ states exists around $E_F$ at the center of the SBZ, but the corresponding band dispersion is unclear. When using 326 eV photons, as for the data shown in Fig. \ref{fig:ARPES-Nb}(i), the dispersion of the $\alpha$ and $\beta$ states are visible, but their intensities are strongly overlapping. Close inspection reveals that there are two separate bands. Green dashed lines in Fig. \ref{fig:ARPES-Nb}(i,j) provide a guide to the eye of the $\alpha$ band dispersion. For 340 eV photons as in Fig. \ref{fig:ARPES-Nb}(j), the intensity of the $\beta$ states becomes relatively weak and the dispersion of the $\alpha$ states becomes visible from $E_F$ to about 0.2 eV below $E_F$. The $\alpha$ band rises from just below $E_F$ at $h\nu=326$ eV to cross $E_F$ at $h\nu=340$ eV. In Fig. \ref{fig:ARPES-Nb}(h), the intensity of the $\alpha$ band is then seen descending from $E_F$ between $h\nu=340$ eV and $h\nu=350$ eV, where the latter corresponds to a BBZ $\Gamma$ point.

As explained in Appendix~\ref{sec:App-A}, the reduced out-of-plane momentum probed at a given photon energy and $k_{||}$ depends on the inner potential at the crystal surface and the $c$-axis lattice constant, which are not the same for the $x=0$ and $x=0.18$ samples. For a clear comparison of the band dispersions exhibited by both samples, we compare cases where the $\delta$ states are cleanly resolved near a $\Gamma$ point of the BBZ. Fig.~\ref{fig:ARPES-Nb}(k) shows the 2D curvature of the ARPES intensity in Fig. \ref{fig:ARPES-hn}(e) for the $x=0$ sample alongside Fig.~\ref{fig:ARPES-Nb}(l), which shows the 2D curvature taken from the intensity in Fig. \ref{fig:ARPES-Nb}(e) for the $x=0.18$ sample. Referring back to Fig. \ref{fig:ARPES-hn}(a-d) and Fig. \ref{fig:ARPES-Nb}(e) shows that each of these measurements coincide with the $\Gamma\mathrm{X}$-line of the BBZ. The 2D curvatures shown were taken from the $k_x$-symmetrized ARPES intensity. The dispersion of the $x=0$ sample $\delta$ band was then found by finding the local extremum of the 2D curvature in each binding-energy slice. This dispersion is plotted over the positive-$k_x$ half of Fig. \ref{fig:ARPES-Nb}(k) with square blue markers. It is shown again with a +0.2 eV shift in Fig. \ref{fig:ARPES-Nb}(l) with square red markers, which overlaps the curvature of the $\delta$ state imaged from the $x=0.18$ sample. This provides additional confirmation that Nb-substitution of Mo at 18\% induces a {quasi-rigid} band shift of $\approx 0.2$~eV.    

\subsubsection{Response to Surface Alkali Deposition}

\begin{figure}
\includegraphics[width=0.60\textwidth]{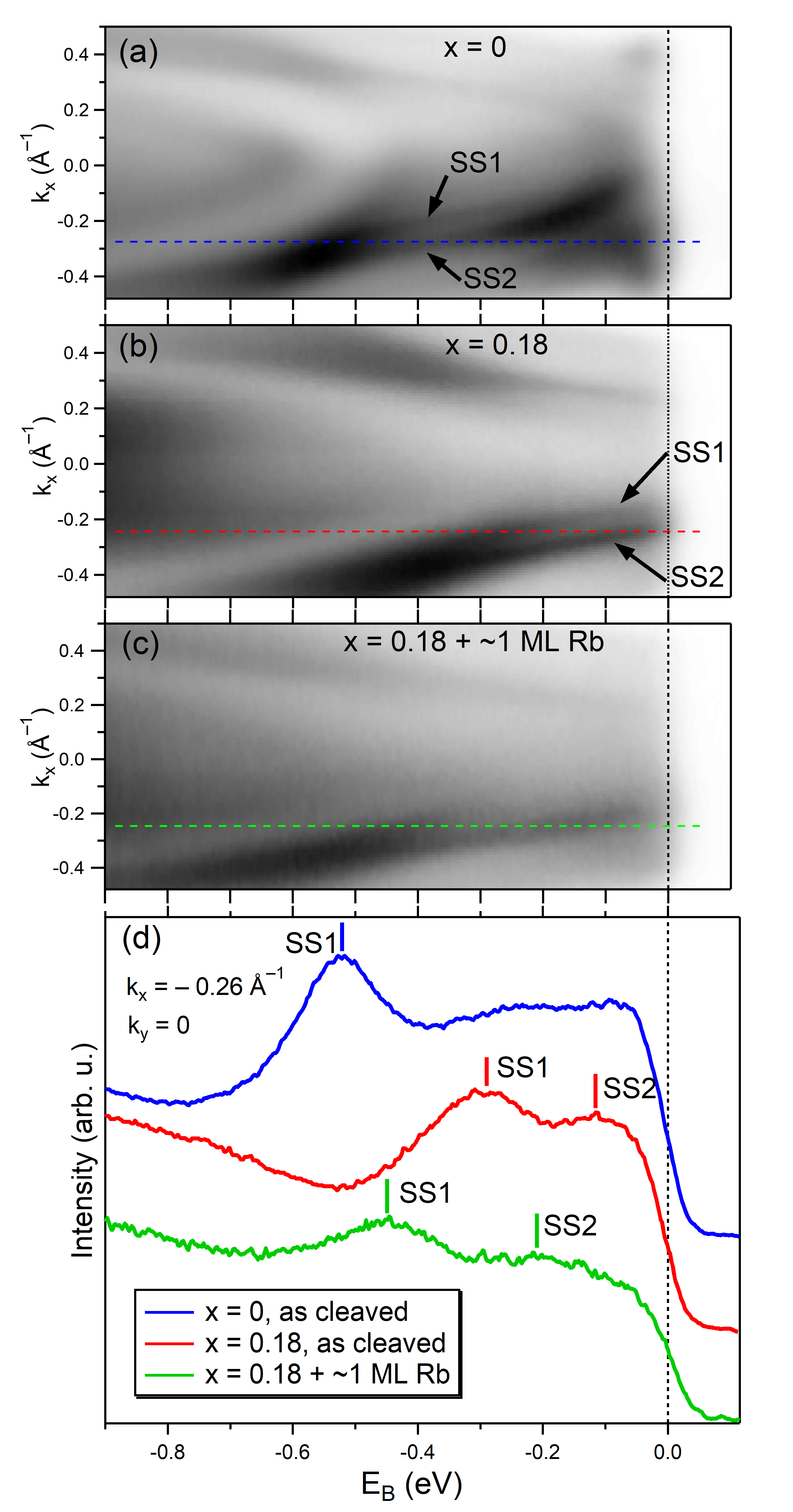} 
\caption{UV-ARPES spectra collected using $p$-polarized 21.22 eV photons with the sample kept at $\approx50$ K. (a-c) Band maps along $\overline{\Gamma X}$ for (a) the $x=0$ and  (b-c) the $x=0.18$ surface (b) before and (c) after the deposition of $\approx 1$ ML of Rb. (d) Energy distribution curves for $k_{y}=-0.26$ \AA$^{-1}$ for each case in (a-c).
}
\label{fig:ARPES-doping}
\end{figure}

Fig. \ref{fig:ARPES-doping} shows UV-ARPES spectra taken along the $k_y=0$ line of the SBZ which capture the changes in band energies following surface deposition of electropositive Rb atoms.
Spectra of the $x=0$ sample surface are shown in Fig. \ref{fig:ARPES-doping}(a) for comparison with those captured for the $x=0.18$ sample surface before and after deposition of $\approx 1$ monolayer (ML) of Rb in Fig. \ref{fig:ARPES-doping}(b) and \ref{fig:ARPES-doping}(c), respectively. Around $k_{x}=-0.35$ \AA$^{-1}$ intense spectral weight appears at an anticrossing of bands around $E_{B}=-0.55$ eV for both the $x=0$ surface and the Rb covered $x=0.18$ surface. It appears around $E_{B}=-0.4$ eV for the pristine $x=0.18$ surface. Yellow circles are drawn around these regions as guides to the eye. Two hole-like bands, SS1 and SS2 discussed previously, disperse toward $E_F$ from this region. For the pristine $x=0.18$ surface, both states reach $E_F$ and maintain a linear dispersion and a momentum splitting of $0.08$ \AA$^{-1}$. For $x=0$, SS2 acquires a larger effective mass starting around $E_B=-0.2$ eV. Both states merge with the bulk continuum around this point and do not reach $E_F$ for the $x=0$ case. Slight mass acquisition is observed in SS2 around this binding energy for the case of Rb covered $x=0.18$ as well.

Fig. \ref{fig:ARPES-doping}(d) shows EDCs of the three spectra for $k_y=-0.26$ \AA$^{-1}$ (marked by dashed lines). The bands are significantly shifted by the alkali coverage with, for example, SS1 shifting 0.15 eV downward in Fig. \ref{fig:ARPES-doping}(d). Spectral weight also appears at $E_F$ around $k_x=-0.4$ \AA$^{-1}$, where $\alpha$ is to appear. It is unclear if this signal instead originates from the $\beta$ states and/or the surface Fermi arc. Spectral weight becomes weaker and broadened due to scattering on the disordered alkali layer, possibly combined with an incoherent surface polar instability, like that experienced in pure 1T'-MoTe$_2$ near room temperature \cite{Weber2018}. Altogether, these effects may preclude clear imaging of states near $E_F$. 
Overall, we interpret the Rb-deposition data as a qualitative indication that alkali adsorption tunes the surface band filling, while the broadened post-deposition spectra and the ambiguity of spectral weight near $E_F$ do not yet allow a precise determination of the doped surface electronic structure.
With a lower degree of Nb-substitution, drastic Lifshitz transitions could be achieved through external means, such as electrostatic or electrolytic doping.

\section{Summary and Conclusion}
\label{sec:summary}

In this work, we have performed a joint theoretical and experimental analysis on the 
Fermi surface evolution of MoTe$_2$ for varying chemical potential, which can be achieved by
Nb-Mo substitution. 
Soft X-ray angle-resolved photoemission spectroscopy results show that Nb substitution significantly reduces the occupation of electronic bands, 
generating isotropic hole pockets  while removing electron pockets in the Fermi surface. 
DFT calculations showed that conduction Bloch states possess a rich and varying orbital texture as a function of chemical potential, where spin-orbit coupling plays an important role. 
Up to eight different bands become involved in the metallic properties of the material for chemical potential levels of the order of 100 meV.

Examining the linear dichroism in ARPES spectra as a function of photon energy and light-polarization, 
indicates both the mixed orbital character of the band structure and atomic-layer-dependent phase shifts that lead to wave-vector-dependent interference in photoemission.
Our discussion of T$_d$-MoTe$_2$ symmetries in Fig.~\ref{fig:Td_struct} described how non-symmorphic symmetry leads to a $4\pi/c$-periodicity in ARPES intensity along $k_z$, which has been observed in previous experiments{~\cite{Pescia1985,Xu2018,PhysRevB.97.235122,PhysRevLett.119.026403,PhysRevX.7.041067}}.
As discussed further in Appendix~\ref{sec:App-A}, an equivalent way to view the origin of this effect is in terms layer-dependent photoemission,  where spatial phase differences between photoelectrons emitted from different atomic layers lead to constructive or destructive interference depending on the final state wave-vector.
The importance of layer-dependent photoemission was recently discussed in the context of T$_d$-WTe$_2$ \cite{Heider2023}.
It must be the case that persistence of $4\pi/c$-periodicity in ARPES intensity we observed in the Nb-doped sample is caused by effectively the same layer-dependent photoemission interference effects, even though the glide-reflection symmetry of the T$_d$ phase is absent and, strictly speaking, no crystalline symmetries exist in the alloy.
The $4\pi/c$-periodicity of ARPES intensity is therefore a necessary, but insufficient, characterization of the T$_d$ phase.
Moreover, the observation of \(4\pi/c\) periodicity with circular polarization and the suppression, rather than redistribution, of photoemission intensity near the Fermi level upon switching from \(p\)- to \(s\)-polarized light in T$_d$-MoTe$_2$ supports the strong \(p_z\) orbital character predicted by DFT.
Models of $1T'$- or T$_d$- MoTe$_2$ that exclude $p_z$ orbitals, such as that of Ref.~\cite{Hu2021}, may struggle to faithfully represent Bloch wave functions in this class of materials.

Our transport results shown in Appendix B also confirm the absence of the transition to the $T_d$ structural phase, a reduced conduction anisotropy,
and reduced magnetoresistance upon $18\%$ Nb substitution. Overall, Nb-Mo partial substitution is a practical means of tuning the rich electronic properties of MoTe$_2$ while maintaining coherent band structure.
The demonstration of Rb surface doping on a bulk $1T'$-Mo$_{0.82}$Nb$_{0.18}$Te$_2$ crystal further suggests that electrostatic gating and/or interface engineering could additionally be used to achieve changes in band filling, potentially driving Lifshitz transitions and fine-tuning electronic properties.


As an outlook, our work shows that Nb substitution provides an efficient means of tuning the low-energy Fermi surface of MoTe$_2$, a capability that may be relevant to its proposed type-II Weyl semimetal phase~\cite{Soluyanov2015,Deng2016,Huang2016}. Since the Weyl nodes in MoTe$_2$ are predicted to lie close to the Fermi level, the observed changes in chemical potential and band dispersions could influence their relative energetic and momentum-space positions, although this lies beyond the scope of the present study. Likewise, because $T_d$-MoTe$_2$ exhibits superconductivity under pressure or chemical tuning~\cite{PhysRevB.94.054517,Guguchia2017}, the demonstrated Fermi-surface control may provide a useful platform for future investigations of the interplay between the electronic structure and superconductivity. Exploring these possibilities will be an interesting direction for future work.

\section*{Acknowledgements}
The authors thank Helmuth Berger for his assistance.
This work was supported by the European Union's Horizon 2020 research and innovation program under the European Research Council (ERC) grant agreement No 946629, the Spanish
MICIU/AEI /10.13039/50110001103 grant, Swiss National Science Foundation Project No. PP00P2\textunderscore144742, No. 200021-137783, No. PP00P2\textunderscore170591, NCCR-MARVEL and the Sino-Swiss Science and Technology Cooperation (Grant No. IZLCZ2-170075). 
M. I. and J. E. O. are funded by the Basque Government Grant IT-2133-26.
P. R. gratefully acknowledges financial support from the DFG (SPP-1666, Project No. MA 4637/3-1) and from the VITI project of the Helmholtz Association, as well as computational support from the JARA-HPC Supercomputing Centre at the RWTH Aachen University.

\section*{Data Availability}
The data that support the findings of this manuscript are
openly available~\cite{ibanez_azpiroz_2026_23390453}.

\appendix
\section{ANALYSIS AND INTERPRETATION OF ARPES DATA}
\label{sec:App-A}
\subsubsection{Symmetry Analysis of ARPES Matrix Element Effects}
Following on the discussion in Sec. \ref{sec:symmetry}C, the matrix element $M_{fi}$ of a dipole-allowed optical transition is written in terms of the dipole operator $\vec{A}\cdot \hat{p}$ as 
$M_{fi} = \bra{f} \vec{A} \cdot \hat{p} \ket{i}$.
The photoelectron current $I$ produced by the transition is proportional to the transition probability $|M_{fi}|^2$. As observed throughout Sec. \ref{subsec:exp}, bands can be hidden or revealed in ARPES spectra according the their symmetry as $\vec{A}$ is varied. Selection rules for a given transition are relatively simple to determine when the dipole operator, the initial state, and final state all have well-defined mirror eigenvalues, as discussed in Refs.~\cite{bassani_band_1967,Moser2017}. For example, we can consider the case that the initial state is in the $M_x$-invariant, $k_x=0$ plane of the BBZ and $\vec{A}$ is aligned to the $x$- or $y$-axis. If the non-relativistic components of both the initial and final states are of even parity under $M_x$, then the transition is forbidden when $\vec{A}$ is aligned to the $y$-axis. If the initial state is odd and the final state remains even, the transition is forbidden when $\vec{A}$ is aligned to the $x$-axis.
It is often assumed that the final state is a free electron plane wave with even parity \cite{Aryal2019}, but in the case of a non-symmorphic space group, the parity of the final state depends on the length of its wave-vector \cite{Pescia1985,Prince1986,landolt2013}.
This, furthermore, depends on the photon energy used \cite{Xu2018}, as described in the following section of this appendix.

Previous studies of T$_d$-MoTe$_2$ \cite{Aryal2019,Ono2021} provided interpretations of the photoemission spectra based on the projected densities of Te $5p$ and Mo $4d$ atomic orbital character in the Bloch wave functions. Our corresponding calculations are shown above in Fig. \ref{fig:DFT-surface-bands}(b-i). Ref. \cite{Aryal2019} employed a free electron final state approximation (FEFSA) wherein the final state wave function $\braket{\vec{r}|\vec{k}_{||}}$ was simply $\propto e^{ik_xx+ik_yy}$ with no dependence on $z$. In this case, the possibility that $M_{fi}\neq0$ for each orbital can be easily worked out from symmetry considerations for cases where $\vec{A}$ is aligned perpendicular or parallel to the surface. Ref. \cite{Ono2021} considered selection rules dictated by the conservation of angular momentum between initial and final states represented as atomic orbitals. They focused on the $l\rightarrow l-1$ channel for the cases of $\vec{A}$ aligned to $\hat{x},\hat{y}$, and $\hat{z}$. This analysis compared well with state-of-the-art one-step photoemission calculations for photon energies between 6 and 60 eV, although a $4\pi/c$ modulation in the ARPES intensity, due to the non-symmorphic symmetry as previously explained, was also observed. It was mentioned that the more complicated, but less restrictive, $l\rightarrow l+1$ excitation channel will begin to dominate at energies (i.e. SX-ARPES). The FEFSA could be thought of as a limiting case as the photon energy, and therefore the kinetic energy of the final state, increases. Even if one insists on expressing the final state in terms of spherical waves centered around each of the atoms in the lattice, from the correspondence principle $(\hbar \rightarrow0)$ in the limit of large quantum numbers it is known that energy quantization vanishes. Therefore, as the photon energy increases, a plane wave (whose expansion as a linear combination of spherical waves is well-known) eventually serves as a good approximation for the final state. The FEFSA as applied in Ref.~\cite{Aryal2019} assumes that the final state wave-function is uniform in the direction perpendicular to the surface.

The results of the selection rule analysis in Refs. \cite{Aryal2019} and \cite{Ono2021} are displayed in Table \ref{table:1}. Both approaches are not equipped to explain photon energy dependent effects, but we can use them to relate strong, qualitative differences in the light-polarization dependent spectra that are consistent across a broad range of photon energies with the orbital projected densities in Fig. \ref{fig:DFT-surface-bands}.
One aspect common to both frameworks is $M_{fi}\neq0$ for excitation from a $p_z$ initial state when $\vec{A}$ is parallel to $\hat{z}$. Our orbital-projected density plots in Fig. \ref{fig:DFT-surface-bands} show that the continuum of $\gamma$ states have a high projected density to the $p_z$ orbitals and very little else. This explains why the $\gamma$ states are not visible when $s$-polarized light is used. Similarly, in the comparison of Fig. \ref{fig:ARPES-hn}(c) and (d), the intensity contributed by the $\alpha$ states reduced drastically with $s$-polarized light across the entire photon energy range studied. Looking back to Fig. \ref{fig:DFT-surface-bands}, the continuum of $\alpha$ states contains $p_z$, $d_{z^2}$, and $d_{yz}$ character, with the latter being focused toward the energy minima of the continuum $(\alpha^*)$. If we subscribe to the $l\rightarrow l-1$ rule, the presence of intensity in that region under $s$-polarized light in Fig. \ref{fig:ARPES-bands}(c) can be explained because $\vec{A}$ was parallel to $\hat{y}$ in the geometry used in that measurement $(\theta_3=0)$ such that the excitation from a $d_{yz}$ initial state is allowed. The data in Fig. \ref{fig:ARPES-hn} were collected with $\theta_3=90^{\circ}$, such that $\vec{A}$ was parallel to $\hat{x}$, such that the same excitation is forbidden. Meanwhile, the vanishing intensity of the $\epsilon$ states under $s$-polarized light in comparing Fig. \ref{fig:ARPES-hn}(e) and (g) is consistent with their strong $p_z$-projected density shown in Fig. \ref{fig:DFT-surface-bands}(d). In addition, the $\beta$ states have visible projected density in all of the orbitals considered and the $\delta$ states have visible $p_x$, $p_y$, $p_z$, $d_{xy}$, and $d_{x^2-y^2}$ projections within our region of interest (from 0.1 to 0.4 eV below $E_F$). This corresponds to the observation of these states with $s$- or $p$-polarized light.

\subsubsection{Transformation Procedures for SX-ARPES Spectra}
The detector collected images of photoemission current as a function of kinetic energy $E_K$ and electron emission angle $\theta_{em}$ within the plane of light-incidence (the $x'z'$-plane referring back to Fig. \ref{fig:ARPES-bands}).
Given a sample tilt rotation $\theta_1$, angle of light incidence $\theta_i$, inner potential barrier $V_0$, and $E_K=h\nu-\phi-E_B$, where $\phi$ is the detector work function, the calculation of the initial state wave-vector components $k_z$ and $k_{||}$ perpendicular and parallel to the sample surface follow as: 
%
%
%
%

%
\begin{eqnarray}\label{eq:kpar}
k_{||}=\dfrac{\sqrt{2m(h\nu-\phi-E_B)}}{\hbar} \sin(\theta_{em}+\theta_1) - h\nu/{\hbar c} \sin(\theta_i+\theta_1),
\end{eqnarray}
\begin{eqnarray}\label{eq:kz}
k_{z}=\sqrt{2m(h\nu-\phi-E_B+V_0)/\hbar^2-k_{||}^2} + h\nu/{\hbar c_{light}} \cos(\theta_i+\theta_1).
\end{eqnarray}
where the photon momentum $h\nu/c_{light}$ contributes a wave-vector shift and $c_{light}$ is the speed of light.
Values of $V_0$ were determined experimentally and independently for the $x=0$ and $x=0.18$ samples by examining the periodicity of spectral features with photon energy.
For each sample, a sequence of photon energies corresponding to the Brillouin zone centers ($\Gamma$ points) were identified from the local minima of the $\alpha$ states. 
$V_0$ was treated as a free parameter and determined by least-squares minimization of the calculated $k_z$ values and the ideal lattice grid of $\Gamma$ points along $k_z$ (integer multiples of $2 \pi/c$).
This procedure yielded $V_0=11.3$ eV and $V_0=12.9$ eV for the $x=0$ and $x=0.18$ samples, respectively.

\section{MAGNETOTRANSPORT MEASUREMENTS}
\label{sec:App-B}
Magnetotransport data were collected with the four-point probe method in a Quantum Design Physical Properties Measurement System at temperatures $2\leq T \leq 300$ K and magnetic fields $0\leq B \leq 9$ T. The results are shown below in Fig. \ref{fig:PPMS}.

Fig. \ref{fig:PPMS}(a) and Fig. \ref{fig:PPMS}(b) show the temperature dependent resistance of $x=0$ and $x=0.18$ crystals. The resistance along the $x$- and $y$-direction, $R_{xx}$ and $R_{yy}$ are shown. 
As in the main text, we define the $xy$-plane as lying parallel to the MoTe$_2$ trilayers, with $\hat{x}$ parallel to the Mo-Mo zigzag chain direction. 
In Fig. \ref{fig:PPMS}(a) the thermal hysteresis anomaly clearly appears in $R_{yy}$ around the transition temperature $T_S\approx 250$ K. 
This corresponds to the first-order transition between the higher-resistance, centrosymmetric 1T' and the lower-resistance, noncentrosymmetric T$_d$ crystalline structure \cite{Qi2016}. 
This feature does not appear for the $x=0.18$ sample in Fig. 1(b), consistent with the absence of the transition reported in ref. \cite{Sakai2016} for this case. 
Resistance anisotropy is apparent for both materials that varies slightly with temperature. 
This is more so in the pure MoTe$_2$ where $R_{yy}/R_{xx}$ is around 3.5 at 300 K and around 2 below 100 K. 
For the $x=0.18$ sample, $R_{yy}/R_{xx}$ remains close to 3.8. 
The residual resistivity ratio (RRR) defined as $R_{yy}(300 K)/R_{yy}(2 K)$ is around 25 and 2 for the $x=0$ and $0.2$ samples, respectively. 
The reduction of RRR is consistent with increased crystalline disorder. 
In this respect, our pure sample is of comparable quality to that studied in ref. \cite{Qi2016}.

Fig. \ref{fig:PPMS}(c) and (d) show Hall resistivity $\rho_{xy}(B)$ curves for temperatures of 300, 200, 50, 10, 5, and 2 K, collected in that order. The sign of the $\rho_{xy}$ is negative for the pure sample and positive for the $x=0.18$ sample, demonstrating that predominately electron-like and hole-like charge carriers exist in the two respective cases. For the $x=0.18$ sample, the $\rho_{xy}(B)$ curve remains nearly linear. 
In the pure sample, the magnitude of $\rho_{xy}(B)$ increases dramatically upon cooling from 300 to 200 K. 
This increase likely results from occupation of the bulk electron-like states that occurs during the transition to the T$_d$ phase \cite{Weber2018}. At 200K $\rho_{xy}$ is linear with $B$. 
At 50 K, a non-linear contribution to $\rho_{xy}(B)$ begins to show, with a downward inflection in the curve occurring around $B=6$ T. 
This component grows as the temperature decreases and saturates; the curves at 10, 5, and 2 K are overlapping. Fig. \ref{fig:PPMS}(e) and (f) show the corresponding magnetoresistance curves $MR(B)\equiv[R_{yy}(B)-R_{yy}(0)]/R_{yy}(0)$. 
The MR reaches 68\% at 9 T for the $x=0$ sample. 
MR in excess of 400\% at 9 T has been reported for MoTe$_2$ \cite{Lee2018}, but MR is known to be very sensitive to the crystal quality/RRR in (Mo/W)Te$_2$ \cite{Cho2017,Flynn2015,Lv2016}. 
Although MR is significantly suppressed in the $x=0.18$ sample to 2.2\% at 9 T, the samples share some trends. 
At 300 and 200 K, the MR is very weak throughout the range of magnetic field. 
Appreciable MR appears at 50 K for fields above 3 T. 
Below 50 K, the MR further increases and the curves are essentially overlapping at the temperatures of 10, 5, and 2 K. 
Aside from the differences in MR magnitude, the profiles of $MR(B)$ at these lower temperatures differ in shape. 
For $B>4$ T, the MR increases more (less) rapidly with $B$ for the $x=0$ ($x=0.18$) case. 
It appears that MR will eventually saturate in the $x=0.18$ case, unlike the pure sample \cite{Lee2018}. 
Much of the reduction in MR is likely related to disorder of the Nb substitution, noting that trace concentrations of aliovalent transition metals in WTe$_2$ reduce both the RRR and MR by an order of magnitude \cite{Flynn2015}.

In addition, the increase in Te deficiency from $x=0$ to $x=0.18$ can also play a role. Studies of non-stoichiometric WTe$_2$ show that chalcogen vacancies reduce magnetoresistance through the combined effects of carrier imbalance and reduced mobility~\cite{Gon17,Yang-Yang2017}. Therefore, the strong reduction in RRR and MR 
can in general be interpreted as  a combined effect of Nb alloying, increased Te deficiency, and disorder.

\newpage
\section*{References}
\footnotesize
\bibliography{References}
\bibliographystyle{apsrev4-1}
 
\expandafter\ifx\csname url\endcsname\relax
 \def\url#1{\texttt{#1}}\fi
\expandafter\ifx\csname urlprefix\endcsname\relax\def\urlprefix{URL }\fi
\providecommand{\bibinfo}[2]{#2}
\providecommand{\eprint}[2][]{\url{#2}}
\clearpage
\clearpage
\clearpage

\clearpage
\clearpage
\clearpage
\clearpage
\clearpage
\clearpage
\clearpage
\clearpage
\begin{table}[h!]
\centering
\begin{tabular}{|p{1cm}||p{1cm}|p{1cm}|p{1cm}|p{1.1cm}|p{1.1cm}|}
 \hline
 \multicolumn{1}{|c|}{} &
 \multicolumn{3}{|c|}{$l\rightarrow l-1$ Rule} & 
 \multicolumn{2}{|c|}{FEFSA}\\
 \hline
 $\braket{ \vec{r } | i }$ & 
 $\vec{A}\parallel\hat{x}$ &
 $\vec{A}\parallel\hat{y}$ &
 $\vec{A}\parallel\hat{z}$ & 
 $\vec{A}\perp\hat{z}$ &
 $\vec{A}\parallel\hat{z}$ \\
 \hline
 $p_x$   & $s$ &  & & $\braket{ \vec{r } | \vec{k}_{||} }$ &\\
 \hline
 $p_y$   &  & $s$ & & $\braket{ \vec{r } | \vec{k}_{||} }$ &\\
 \hline
 $p_z$   &  & & $s$ & & $\braket{ \vec{r } | \vec{k}_{||} }$\\
 \hline
 $d_{xz}$  & $p_z$ & & & & $\braket{ \vec{r } | \vec{k}_{||} }$\\
 \hline
 $d_{yz}$  & & $p_z$ & & & $\braket{ \vec{r } | \vec{k}_{||} }$\\
 \hline
 $d_{z^2}$ & & & $p_z$ & $\braket{ \vec{r } | \vec{k}_{||} }$ &\\
 \hline
 $d_{xy}$  & & & $p_x$ & $\braket{ \vec{r } | \vec{k}_{||} }$ &\\
 \hline
 $d_{x^2-y^2}
 $ & & & & $\braket{ \vec{r} | \vec{k}_{||} }$ &\\
 \hline
\end{tabular}
\caption{
Summary of the dipole-allowed transitions for the $l\rightarrow l-1$ selection rule discussed in Ref. \cite{Ono2021} and the free electron plane wave final state approximation (FEFSA) used in Ref. \cite{Aryal2019} for given orientations of the vector potential. The leftmost column lists the initial states and the remaining columns show the corresponding final states for dipole-allowed transitions. Blank entries indicate that any transition is dipole-forbidden.
}
\label{table:1}
\end{table}

\clearpage
\begin{figure}
\includegraphics[width=0.99\textwidth]{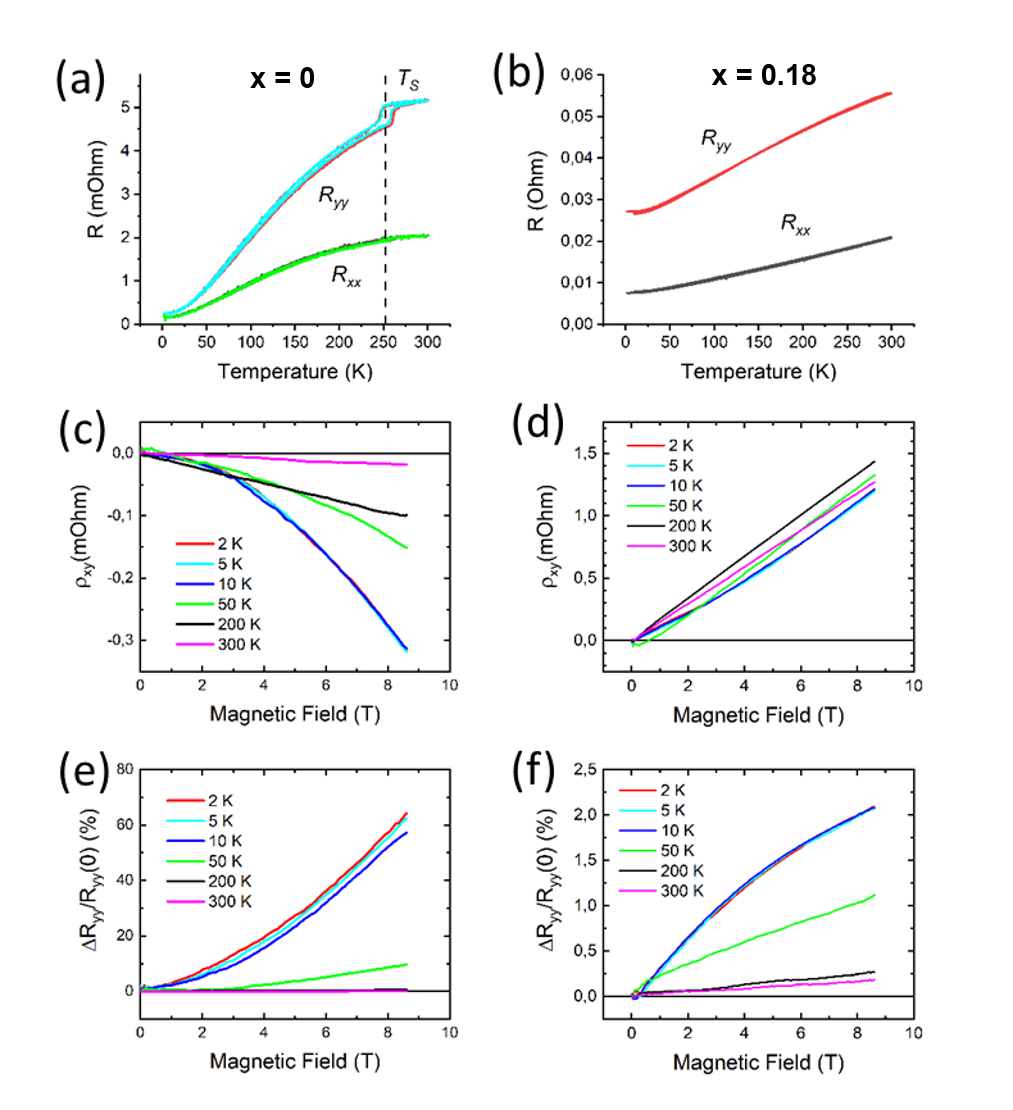}
\caption{Transport data for (a,c,e) $x = 0$ and (b,d,f) $x=0.18$ samples. (a-b) Temperature-dependent resistance along the $x$-direction ($R_{xx}$) and $y$-direction ($R_{yy}$). (c-d) Magnetic-field-dependent Hall resistivity and (e-f) magnetoresistance collected at various temperatures.}
\label{fig:PPMS}
\end{figure}

\end{document}